\documentclass[10pt,conference]{IEEEtran}
\IEEEoverridecommandlockouts
\usepackage{cite}
\usepackage{amsmath,amssymb,amsfonts}
\usepackage{algorithmic}
\usepackage{graphicx}
\usepackage{textcomp}
\usepackage{xcolor}
\usepackage{subcaption}
\usepackage{multirow}
\usepackage{tablefootnote}
\usepackage[framemethod=tikz]{mdframed}
\usepackage{adjustbox}
\usepackage{soul}

\usepackage{soul} 

\def\BibTeX{{\rm B\kern-.05em{\sc i\kern-.025em b}\kern-.08em
    T\kern-.1667em\lower.7ex\hbox{E}\kern-.125emX}}
\begin{document}

\title{An Empirical Study on Automatically Detecting AI-Generated Source Code: How Far Are We?}

\makeatletter
\newcommand{\linebreakand}{%
  \end{@IEEEauthorhalign}
  \hfill\mbox{}\par
  \mbox{}\hfill\begin{@IEEEauthorhalign}
}
\makeatother

\author{\IEEEauthorblockN{Hyunjae Suh}
\IEEEauthorblockA{
\textit{University of California, Irvine}\\
Irvine, CA, USA \\
hyunjas@uci.edu}
\and
\IEEEauthorblockN{Mahan Tafreshipour}
\IEEEauthorblockA{
\textit{University of California, Irvine}\\
Irvine, CA, USA \\
mtafresh@uci.edu}
\and
\IEEEauthorblockN{Jiawei Li}
\IEEEauthorblockA{
\textit{University of California, Irvine}\\
Irvine, CA, USA \\
jiawl28@uci.edu}
\linebreakand
\IEEEauthorblockN{Adithya Bhattiprolu}
\IEEEauthorblockA{
\textit{University of California, Irvine}\\
Irvine, CA, USA \\
abhattip@uci.edu}
\and
\IEEEauthorblockN{Iftekhar Ahmed}
\IEEEauthorblockA{
\textit{University of California, Irvine}\\
Irvine, CA, USA \\
iftekha@uci.edu}
}


\maketitle

\begin{abstract}
Artificial Intelligence (AI) techniques, especially Large Language Models (LLMs), have started gaining popularity among researchers and software developers for generating source code. However, LLMs have been shown to generate code with quality issues and also incurred copyright/licensing infringements. Therefore, detecting whether a piece of source code is written by humans or AI has become necessary. This study first presents an empirical analysis to investigate the effectiveness of the existing AI detection tools in detecting AI-generated code. The results show that they all perform poorly and lack sufficient generalizability to be practically deployed. Then, to improve the performance of AI-generated code detection, we propose a range of approaches, including fine-tuning the LLMs and machine learning-based classification with static code metrics or code embedding generated from Abstract Syntax Tree (AST). Our best model outperforms state-of-the-art AI-generated code detector (GPTSniffer) and achieves an F1 score of 82.55. We also conduct an ablation study on our best-performing model to investigate the impact of different source code features on its performance.
\end{abstract}


\section{Introduction}
\label{sec:intro}


Artificial Intelligence (AI), such as machine learning techniques, has been widely used to tackle software development tasks, especially for the generation of source code~\cite{shim2020deepercoder, lu2021codexglue, dehaerne2022code, siddiq2022securityeval}. More recently, Large Language Models (LLMs) that were pre-trained on large and diverse data corpora have shown state-of-the-art performance in code generation~\cite{chen2021evaluating, li2022competition, luo2023wizardcoder, ren2023misuse, liu2023codegen4libs, yu2024codereval}. The generative LLMs such as ChatGPT~\cite{chatgpt}, Gemini Pro~\cite{gemini} and Starcoder2~\cite{lozhkov2024starcoder2stackv2} are able to generate code that is very similar to what a human developer would produce given the natural language specification. While a large amount of previous research works have explored a variety of fine-tuning/prompting techniques~\cite{radford2018improving, liu2023pre} to further boost model's performance in generating high-quality source code, numerous LLM-based tools (i.e., Github Copilot~\cite{copilot}) have been implemented to assist developers to design software architecture, generate production code/test cases, and refactor the existing code base. Therefore, leveraging LLMs for source code generation or assisting programming-related tasks is becoming popular among software practitioners. 

However, the wide adoption of LLMs for code generation has raised a variety of concerns among researchers and software practitioners. Researchers have questioned the evaluation process of the source code quality generated by LLMs~\cite{liu2024your}, and the correctnesss of the generated code could be easily impacted by the wording in the LLMs' prompts~\cite{mastropaolo2023robustness}. In addition, it's been proven that around 35\% of Github Copilot-generated code snippets on Github have security issues of various types~\cite{fu2023security,asare2023github}, indicating the security risks of the code generated by LLMs. Moreover, the violations of intellectual property rights have also been found in LLMs such as generation of licensed code~\cite{yu2023codeipprompt}.


Therefore, it has become necessary to determine whether a code snippet is written by humans or generated by the LLMs. While there exist a variety of automated tools (i.e., GPTZero~\cite{gptzero}, Sapling~\cite{sapling}, and more) to detect Artificial Intelligence Generated Content (AIGC), such tools were built for detecting natural language texts and their performance in detecting AI-generated source code still remains far from being perfect~\cite{nguyen2023snippet, pan2024assessing}. To fill this gap, Nguyen et al.~\cite{nguyen2023snippet} proposed GPTSniffer by fine-tuning CodeBERT~\cite{feng2020codebert} to classify a code snippet as either human-written or LLM-generated. However, they only considered the code that is written in Java programming language and is generated only by ChatGPT. It has been proven that different LLMs are good at generating code for different sets of coding problems~\cite{yu2024codereval}; namely, they tend to perform differently given the same set of coding tasks. Thus, GPTSniffer's generalizability to code that's written in other programming languages or generated by LLMs other than ChatGPT is not investigated. A better detection approach for detecting AI-generated source code is still missing.


In this paper, we first conduct a comprehensive empirical study to evaluate the performance of existing AIGC detectors for detecting AI-generated source code. The goal was twofold: first, as a complement to prior studies~\cite{nguyen2023snippet, pan2024assessing}, we investigate the widely-adopted AIGC detectors' ability to detect AI-generated source code that is written in various programming languages, generated by multiple popular generative LLMs, and from different domains (i.e., programming questions, open-source development tasks). Second, we analyzed the performance of the current state-of-the-art detector specifically for source code, GPTSniffer. Therefore, we asked the following research questions:

\textbf{RQ1: How do existing AIGC detectors perform on detecting AI-generated source code?}




\textbf{RQ2: How can we improve the performance of AI-generated source code detection?}


\textbf{RQ3: How do the source code features captured by embeddings contribute to the overall effectiveness?}


The significance of our contributions are following:

\begin{itemize}
\item We show that existing AIGC detectors for text perform poorly in detecting AI-generated source code.
\item We show that the current state-of-the-art technique for AI-generated code detection, GPTSniffer, fails to generalize effectively across different programming languages, programming tasks, and generative LLMs.
\item We built a variety of machine learning and LLM-based classifiers to detect AI-generated code, which outperform the compared techniques and show decent performance across multiple programming languages, programming tasks, and generative LLMs.
\end{itemize}

The remainder of this paper is organized as follows: in Section \ref{sec:rw}, we provide related works of AIGC detection and background about LLM-based code generation and pre-trained source code embeddings. We outlined our data collection, model building, and performance analysis in Section \ref{sec:method}. Next, we present the evaluation results and observations in Section \ref{sec:results}. Then, we discuss the implication for our study in Section \ref{sec:discussion}. Section \ref{sec:ttv} shows potential threats to the validity of our approaches and findings. Finally, we conclude with a summary of the findings in Section \ref{sec:conclusion}.
\section{Related Work \& Background}
\label{sec:rw}

\subsection{Large Language Models for Code Generation}
Due to advancements in the field of Natural Language Processing (NLP), LLMs have seen substantial progress in their performance and widespread use~\cite{brown2020language}. More recently, source code has also been included to train the LLMs with the goal of helping with software development activities~\cite{feng2020codebert}. Since these models such as CodeBERT~\cite{feng2020codebert}, CodeT5~\cite{wang2021codet5}, Starcoder2~\cite{lozhkov2024starcoder2stackv2}, and ChatGPT~\cite{chatgpt} have been trained on vast and diverse datasets of source code and natural language from various domains~\cite{dai2015semi}, they showed cutting-edge effectiveness when being fine-tuned~\cite{radford2018improving} or prompted~\cite{liu2023pre} to solve various downstream Software Engineering (SE) tasks ~\cite{geng2024large,guo2024exploring, feng2024prompting}. 

LLMs are widely adopted to directly generate source code given the docstring/requirement in natural language (code generation) or complete the code for the developers based on the software context (code completion)~\cite{lu2021codexglue}. To further ease the use of LLMs for code generation/code completion and improve the quality of the generated code, researchers have been investigating fine-tuning/prompting approaches to generate code of high quality that meets developers' intentions~\cite{ren2023misuse, liu2023codegen4libs, yu2024codereval}. In addition, numerous software development tools for code generation/code completion, such as Github Copilot~\cite{copilot}, have been released and widely used by developers. All these techniques significantly increase the chance that AI (i.e., LLMs) writes a piece of code instead of a human developer.


In this study, we specifically analyzed the code generated by Gemini Pro~\cite{gemini}, ChatGPT, GPT-4~\cite{gpt-4}, and Starcoder2-Instruct (15B)~\cite{lozhkov2024starcoder2stackv2} since they are among the state-of-the-art LLMs and have been the subjects of many prior SE studies in code generation~\cite{khoury2023secure, ren2023misuse, poldrack2023ai, liu2023codegen4libs, yu2024codereval, liu2024your, rane2024gemini, jiang2024surveylargelanguagemodels}. We believe that this selection of LLMs covers not only the detectability of code generated by general LLMs (i.e., ChatGPT, Gemini Pro, and GPT-4) but also that of code-specific LLMs (i.e., Starcoder2-Instruct).

\subsection{Automated Detection of Artificial Intelligence Generated Content (AIGC)}
The emergence of generative LLMs such as ChatGPT has surged the demand for accurately detecting AIGC. A variety of AIGC detectors have been developed. For example, GPTZero~\cite{gptzero} is a widely-used commercial AIGC detector, while 
Sapling~\cite{sapling} generates the probability of whether each token in the input is AIGC. It reached 97\% accuracy in identifying AI-generated texts. 
In addition, researchers and open-source software practitioners have also been actively developing AIGC detectors such as GPT-2 Detector~\cite{gpt2detector}, DetectGPT~\cite{mitchell2023detectgpt}, and Giant Language Model Test Room (GLTR)~\cite{strobelt2019catching}, which achieved decent performance in detecting AIGC.

However, the aforementioned AIGC detectors are only designed to detect AI-generated natural language texts. Since source code has unique linguistic syntax and writing styles that differ from natural language~\cite{pan2024assessing}, these detectors may not perform well in determining whether a human or AI writes the source code snippet. Researchers have recently found that these text detectors showed limited effectiveness when detecting AI-generated code~\cite{nguyen2023snippet, pan2024assessing}. To fill this gap, Nguyen et al.~\cite{nguyen2023snippet} proposed GPTSniffer, where they fine-tuned CodeBERT to classify whether a code snippet is written by AI or human. However, it could not generalize well to the data that it was not trained on (Section \ref{sec:aigc_detector_result}). In addition, only ChatGPT was queried to generate the code, while Java was the only programming language considered. Since there is a gap in terms of comprehensive analysis across different languages and generative LLMs, in our study, we investigated our approaches and the existing AIGC detectors' performance on the code generated by four widely-used state-of-the-art LLMs, namely, Gemini Pro, ChatGPT, GPT-4, and Starcoder2-Instruct. We also experimented with C++ and Python in addition to Java. In order to ensure the generalizability of our approaches across different datasets, we selected three widely-used code generation benchmarks, MBPP~\cite{austin2021program}, HumanEval-X~\cite{zheng2023codegeex}, and CodeSearchNet~\cite{husain2019codesearchnet}.

\subsection{Pre-trained Source Code Embedding}
Distributed numeric code representations, pre-trained code embeddings, have been proven to be effective in various SE tasks, such as automated program repair~\cite{wang2017dynamic, chen2018remarkable,white2019sorting}, vulnerability prediction~\cite{harer2018automated,pradel2018deepbugs}, and code clone detection~\cite{buch2019learning}. Various pre-trained embedding models have been proposed by researchers to better capture syntactical/semantic information of source code and improve downstream SE tasks~\cite{zhang2019novel, ding2022can}. 

More recently, researchers have started incorporating structural information (i.e., information from Abstract Syntax Tree (AST)) and the textual information of source code into code embeddings. For example, Zhang et al.~\cite{zhang2019novel} first split each large AST into smaller statement ASTs and encoded these ASTs to numeric vectors by capturing the lexical and syntactical knowledge of statements. They then used a bidirectional Recurrent Neural Network (RNN) to leverage the statements' naturalness and produce the embedding. Ding et al.~\cite{ding2022can} used a two-step unsupervised training strategy to integrate the textual and structural information from the code to make the embeddings more generalized to different SE tasks. They found that including structural information could help improve the code embeddings' quality.


Since code embeddings are usually obtained by training models with large source code datasets to acquire knowledge about the semantic and syntactic meaning, we posit machine learning models trained with such embeddings have the potential to perform better than models without such information when trying to differentiate human-written code from AI-generated code. In this study, we selected CodeT5+ 110M embedding model~\cite{wang2023codet5+}, which shows state-of-the-art performance on code understanding and generation tasks, to generate embeddings for source code and AST to incorporate both textual and structural information. Then, we trained our machine learning models with these embeddings to detect AI-generated code with the goal of achieving decent performance.


\section{Methodology}
\label{sec:method}

Our goal was to investigate the effectiveness of the current AIGC detectors in detecting AI-generated source code (RQ1). Our other goal was to classify a code snippet as human-written or AI-generated (RQ2). Finally, we analyze the impact of various source code features on the performance of the best-performing approach (RQ3). In the following subsections, we detail the applied methodology. Figure \ref{fig:overview} shows an overview of our methodology.


\begin{figure}[htp]
    \centering
    \includegraphics[width=0.8\columnwidth]{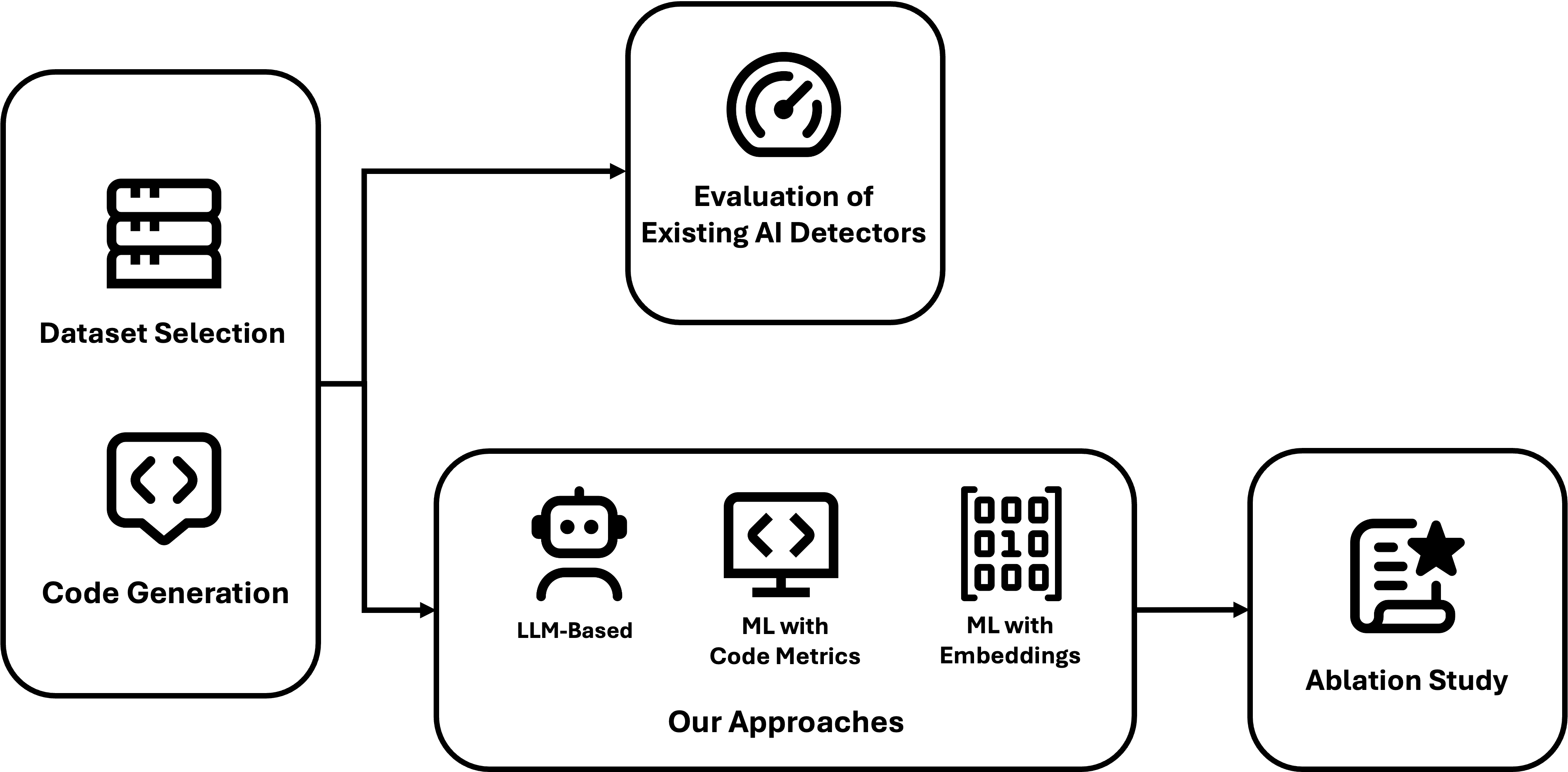}
    \caption{Overview of Research Method}
    \label{fig:overview}
\end{figure}

\subsection{Data Collection}
\label{sec:datacollection}

Since we aimed to detect AI-generated source code and evaluate the effectiveness of the current AIGC detectors, we targeted the code generation benchmark datasets that have been studied by previous research related to code generation~\cite{khoury2023secure, ren2023misuse, poldrack2023ai, liu2023codegen4libs, yu2024codereval, liu2024your, rane2024gemini}. Specifically, we selected three datasets, namely, MBPP~\cite{austin2021program}, HumanEval-X~\cite{zheng2023codegeex}, and CodeSearchNet~\cite{husain2019codesearchnet}. MBPP contains 974 crowd-sourced Python programming problems along with human-written Python functions solving the specified problems. The problems range from simple numeric manipulations to tasks that require basic usage of standard library functions. HumanEval-X consists of 820 data samples (function docstrings/specifications and the corresponding human-written code solutions) in Python, C++, Java, JavaScript, and Go. To make our evaluation more generalized and not restricted to human-crafted coding questions, we also included CodeSearchNet, a dataset of 2 million pairs of comment and human-written code collected from publicly available open-source non-fork GitHub repositories. In this study, we aimed to investigate the generalizability of the AIGC detectors and our approaches across multiple programming languages, so we selected Java, C++, and Python, three most widely-used programming languages by software practitioners~\cite{topprogram,TIOBE}. We believe that selecting only three widely used languages instead of using all the languages available in the studied datasets would keep our experiments manageable. It's worth pointing out that the MBPP dataset only has code written in Python programming language, while CodeSearchNet has Java and Python but does not include C++. To ensure there were no duplicates between different datasets, we manually reviewed all specifications and code snippets. Additionally, we used a clone detection tool, Nicad~\cite{nicad}, to identify any potential code clones. Consequently, we found no duplicates or clones across different datasets.

To collect the AI-generated counterparts for the human-written code in the datasets, we adopted four state-of-the-art generative LLMs that have been widely used in code generation literature, namely, ChatGPT~\cite{chatgpt}, Gemini Pro~\cite{gemini}, GPT-4~\cite{gpt-4}, and Starcoder2-Instruct~\cite{lozhkov2024starcoder2stackv2}, to generate source code based on the natural language specifications or comments in the selected datasets. Due to a limited financial budget and prohibitively expensive OpenAI API~\cite{openai}, instead of generating code for the complete CodeSearchNet dataset, which consists of 457k Python and 497k Java codes, we randomly sampled 400 (confidence level 95\%, margin of error 5\%) data instances each for Python and Java. 

In addition, the generation from the LLMs tends to be non-deterministic and creative when the temperature increases~\cite{peng2023towards,geng2024large}. To make our evaluation more comprehensive, which covers code generations with greater linguistic variety, we generated multiple code snippets for each specification using the same LLM with different temperatures. To make our experiments controllable, we set the temperature as 0 and the default value provided by respective LLMs (For ChatGPT, GPT-4, and Starcoder2-Instruct, the default temperature is 1. For Gemini Pro, the default is 0.9) to generate code based on the specifications. 

After code generation, we obtained the code generated by each selected LLM. Combining the code generated by LLM with the human-written code which already existed in the original dataset, we obtained datasets with twice the size of original one. Then we removed the data instances where the LLMs could not generate the source code given the specifications. We also removed code snippets that contain syntax errors, which would prevent us from extracting static code features later in Section \ref{sec:ml}. The statistics of the collected datasets are shown in Table \ref{tab:datat0} and Table \ref{tab:datat1}.

\begin{table}
\centering
\caption{Collected Dataset with AI-generated Code (Temperature = 0)}
\label{tab:datat0}
\begin{adjustbox}{max width=\textwidth}
\tiny
\begin{tabular}{l|l|l|l|l} 
\hline
                                & \textbf{ChatGPT} & \textbf{Gemini Pro} & \textbf{GPT-4} & \textbf{Starcoder2-Instruct}  \\ 
\hline
\textbf{MBPP (Python)}          & 1,946            & 1,948               & 1,932          & 1,948                \\ 
\hline
\textbf{HumanEval-X (Python)}   & 323              & 327                 & 323            & 328                  \\ 
\hline
\textbf{HumanEval-X (Java)}     & 281              & 320                 & 311            & 328                  \\ 
\hline
\textbf{HumanEval-X (C++)}      & 328              & 325                 & 328            & 328                  \\ 
\hline
\textbf{CodeSearchNet (Python)} & 400              & 400                 & 400            & 393                  \\ 
\hline
\textbf{CodeSearchNet (Java)}   & 397              & 400                 & 384            & 395                  \\
\hline
\end{tabular}
\end{adjustbox}
\end{table}


\begin{table}
\centering
\caption{Collected Dataset with AI-generated Code (Default Temperature)}
\label{tab:datat1}
\begin{adjustbox}{max width=\textwidth}
\tiny
\begin{tabular}{l|l|l|l|l} 
\hline
                                & \textbf{ChatGPT} & \textbf{Gemini Pro} & \textbf{GPT-4} & \textbf{Starcoder2-Instruct}  \\ 
\hline
\textbf{MBPP (Python)}          & 1,933            & 1,948               & 1,942          & 1,948                \\ 
\hline
\textbf{HumanEval-X (Python)}   & 324              & 327                 & 323            & 328                  \\ 
\hline
\textbf{HumanEval-X (Java)}     & 309              & 319                 & 314            & 323                  \\ 
\hline
\textbf{HumanEval-X (C++)}      & 328              & 325                 & 328            & 317                  \\ 
\hline
\textbf{CodeSearchNet (Python)} & 400              & 400                 & 400            & 394                  \\ 
\hline
\textbf{CodeSearchNet (Java)}   & 395              & 383                 & 400            & 390                  \\
\hline
\end{tabular}
\end{adjustbox}
\end{table}

\subsection{Compared AIGC Detectors}
\label{sec:aigc_detector_result}
A number of AIGC detectors have been implemented to detect AI-generated natural language texts. Similar to Pan et al.~\cite{pan2024assessing}, our goal was to investigate the effectiveness of these detectors in detecting AI-generated source code and compare them with our approaches. Following their work, we selected five AIGC detectors: GPTZero~\cite{gptzero}, GPT-2 Output Detector~\cite{gpt2detector}, DetectGPT~\cite{mitchell2023detectgpt}, GLTR~\cite{strobelt2019catching}, and Sapling~\cite{sapling}. We ran these detectors on human-written code from our selected datasets and also on the AI-generated code by the three LLMs with different temperatures. 


More recently, Nguyen et al.~\cite{nguyen2023snippet} built a classifier named GPTSniffer that specifically targeted the identification of AI-generated source code. They fine-tuned CodeBERT~\cite{feng2020codebert} with human-written code and ChatGPT-generated code to classify a code into human-written or AI-generated. In this study, we include GPTSniffer as a baseline to systematically evaluate its performance on AI-generated code from different LLMs, temperatures, and programming languages. 

\subsection{Evaluation Settings and Metrics}
\label{sec:metrics}
Similar to previous works~\cite{nguyen2023snippet}, we split each of the selected dataset using the 80:10:10 ratio, allocating 80\% for training, 10\% for validation, and 10\% for testing. To prevent any data overlap between datasets that are from the same source but generated by different LLMs (i.e. HumanEval-C++-ChatGPT, HumanEval-C++-Gemini Pro, HumanEval-C++-GPT-4), we consistently splitted the datasets. Each of the source code has a ground truth label of either \textit{Human} or \textit{AI}, representing that the code is either generated by humans or the LLMs. The metrics were calculated based on the comparison between the ground truth labels and the predicted labels. The specific metrics we used in this study include \textit{Accuracy}, \textit{True Positive Rate (TPR)}, \textit{True Negative Rate (TNR)}, and \textit{F1-score}. In this study, the positive label stands for \textit{Human}, while the negative label represents \textit{AI}. Detailed explanation of the metrics is as follows:

\textbf{Accuracy:} Accuracy is calculated as the ratio of correct predictions to the total number of predictions. It is calculated as $Accuracy=\frac{TP+TN}{TP+TN+FP+FN}$. TP is the number of human-written code predicted correctly. TN denotes the count of correctly predicted samples of AI-generated code. We defined TN this way since its original definition corresponds to the count of correctly predicted negative class samples, and in our classification scheme, we set \textit{AI} as the negative label. FP is the number of AI-generated code incorrectly predicted as human-written code, and FN is the number of human-written code incorrectly predicted as AI-generated code. 

\textbf{TPR:} True Positive Rate (Recall) represents the ratio of actual positive cases that are correctly identified as positive by the classification model. It is calculated as $TPR=\frac{TP}{TP+FN}$.

\textbf{TNR:} True Negative Rate represents the ratio of actual negative cases that are correctly identified as negative by the classification model. It is calculated as $TNR=\frac{TN}{TN+FP}$.


\textbf{F1-score:} F1 score is the harmonic mean of precision and recall, where precision is calculated as $Precision=\frac{TP}{TP+FP}$. However, it can vary depending on which class (i.e., human-written or AI-generated) is designated as positive, potentially leading to a misrepresentation of the model's performance. For instance, consider a scenario where the \textit{Human} class is designated as positive. If the model accurately predicts all instances of human-generated code but misclassifies all AI-generated code, its F1-score would be zero. Thus, we also calculated two variants of F1-scores by setting either \textit{Human} or \textit{AI} as the positive label. We represented the F1-score where \textit{Human} is set as the positive label as \textit{Human F1-score}, while the F1-score where \textit{AI} is set as the positive label as \textit{AI F1-score}. Finally, with the two variants, we calculated \textit{Average F1-score}, which is the average of the two F1 score variants and is computed by taking the macro average of the \textit{Human F1 score} and \textit{AI F1 score}. It is calculated as $Average\ F1=\frac{Human\ F1+AI\ F1}{2}$. We believe this F1 score can better represent the overall effectiveness.


For each of our proposed techniques (i.e., models) in this study, we conducted the evaluation in two different settings. First, we trained the models on the training split of each of the datasets (Section \ref{sec:datacollection}), conducted hyper-parameter tuning to select the model with the best performance, and evaluated the model's performance on the testing split (\textit{``Within" evaluation setting}). To further evaluate the generalizability of the models across different datasets with programs in different domains and written in different programming languages, we also tested the models on the testing split of another dataset (\textit{``Across" evaluation setting}). For example, we trained a model on MBPP's training data split, and tested it on HumanEval-X's testing split. For the baseline AIGC detectors that we compared ours with, we ran these detectors on the testing splits of our datasets to evaluate their performance.

\subsection{LLM-based Approaches}
\label{sec:LLMbased}
Since LLMs have shown state-of-the-art performance on code classification tasks such as defect detection and clone detection~\cite{niu2023empirical}, we decided to harness the power of LLMs to detect AI-generated code. We used zero-shot learning, in-context learning with retrieved demonstrations, and fine-tuning. In this study, we selected ChatGPT (i.e., GPT-3.5-turbo) as the model to perform prompting/fine-tuning since it is one of the LLMs that showed state-of-the-art performance on a variety of SE tasks~\cite{geng2024large,guo2024exploring,feng2024prompting}. We did not take GPT-4 or Gemini Pro due to the unavailability of model fine-tuning.


In this study, we utilized three different representations of source code 
namely, the textual content of the source code, the AST representation by~\cite{guo2022unixcoder}, and the concatenation of textual code and AST representation. 
Specifically, we followed the algorithm presented by Guo et al.~\cite{guo2022unixcoder} to generate AST representation using Tree-Sitter~\cite{treesitter}, which starts from the root node and recursively traverses the AST, appending node names with special suffixes (i.e., left, right) to the resulting sequence. The models trained with this AST representation showed state-of-the-art performance in various code understanding tasks~\cite{niu2023empirical}. In the following section, we will use \textit{Code Only} for textual code representation, \textit{AST Only} for the AST representation by~\cite{guo2022unixcoder}, and \textit{Combined} for the concatenation of \textit{Code Only} and \textit{AST Only}.


In the zero-shot learning setting, we prompted ChatGPT to determine whether the given source code snippet is generated by AI or human. We followed the established best practices~\cite{awesomeprompt,ibm} to design our prompt (The prompt is provided in the replication package~\cite{replicationpackage}). In the in-context learning setting, we provided demonstration examples from the training datasets in the prompt in addition to the zero-shot setting. Following the best practices of setting in-context learning examples for SE tasks~\cite{gao2023makes}, we used BM-25~\cite{zhang2020retrieval} to retrieve four demonstration examples (two instances of human-written code and two instances of AI-generated code) from the training datasets that are the most similar to the code snippet to be predicted (i.e., test sample), and we ordered the examples based on their similarity to the code snippet in ascending order. In this study, we prompted the model with one of the three code representations to analyze the models' performance using different representations. In the fine-tuning setting, the training data consists of one of the representations (i.e., \textit{Code Only}, \textit{AST Only}, or \textit{Combined}) and the ground truth labels. OpenAI API was called to fine-tune the model (i.e. ChatGPT). In the following sections, we will use \textit{fine-tuned ChatGPT} to represent the ChatGPT model that is fine-tuned to detect AI-generated code with our datasets. We evaluated the models in \textit{``Within" evaluation setting} only for the zero-shot learning setting as it does not require training data. For in-context learning and fine-tuning, we performed evaluation in both \textit{``Within" evaluation setting} and \textit{``Across" evaluation setting}.

\subsection{Machine Learning Classifiers with Static Code Metrics}
\label{sec:ml}
Despite the recent advancement in deep learning and LLMs, shallow machine learning algorithms still showed decent performance in some SE tasks given the appropriate features are provided~\cite{Dias2020Software, alikhashashneh2018using}. In this study, we also investigated the feasibility and effectiveness of a machine learning-based classification approach for AI-generated source code detection. 

To construct the features from source code for training our machine learning models, we used Scitools Understand~\cite{und} to extract static source code metrics as the features since these metrics have been used in a number of previous works~\cite{alikhashashneh2018using, clemente2018predicting, gupta2020extracting, medeiros2020vulnerable, gesi2021empirical}. In addition, we also collected features studied by Aljehane et al.~\cite{aljehane2021determining}, which includes Identifiers (method and variable names), Names and Operators in if, else, and while statements, Operators, Keywords, Arguments, and Method Signatures. We believe that these features also have the potential to be used to distinguish between human-written and AI-generated code since human developers focus on them when they review source code. We used Tree-Sitter~\cite{treesitter} to collect the code features for the three programming languages in our study. Since our datasets consist of three different languages, we filtered the metrics to keep those that are commonly applicable to all three languages. For instance, Python does not have semicolon so we removed all features that are related to semicolon. In total, we retained 30 code features. Due to space constraints, we provided the whole set of these features in our replication package~\cite{replicationpackage}.

In order to avoid multicollinearity across features~\cite{katrutsa2017comprehensive}, we conducted Variance Inflation Factor (VIF) on all 30 features. VIF is a statistical measure used to assess multicollinearity in regression analysis. High VIF values indicate strong multicollinearity, which can lead to unstable and unreliable regression coefficients. Akinwande et al.~\cite{akinwande2015variance} argued that a VIF value between 5 and 10 indicates a high correlation that may be problematic. Thus, we set our threshold value as 5 and make VIF values of our features below that value. 
As a result, we retained eight features. The finalized set of features and their corresponding descriptions are provided in Table \ref{tab:features}. 

\begin{table}
\centering
\caption{Collected Source Code Features}
\label{tab:features}

\begin{adjustbox}{max width=\textwidth}
\begin{tabular}{l|l} 
\hline
\textbf{Features}                                                                      & \textbf{Definitions}                                                                                                                                          \\ 
\hline
SumCyclomatic                                                                          & \begin{tabular}[c]{@{}l@{}}Sum of cyclomatic complexity \\ of all nested functions or methods.\end{tabular}                                              \\ 
\hline
AvgCountLineCode                                                                       & \begin{tabular}[c]{@{}l@{}}Average number of lines contai-\\ -ning source code for all nested\\ functions or methods.\end{tabular}                            \\ 
\hline
CountLineCodeDecl                                                                      & \begin{tabular}[c]{@{}l@{}}Number of lines containing decl-\\ -arative source code\end{tabular}                                                               \\ 
\hline
CountDeclFunction                                                                      & Number of functions                                                                                                                                           \\ 
\hline
MaxNesting                                                                             & \begin{tabular}[c]{@{}l@{}}Maximum nesting level of (if, w-\\ -hile, for, switch etc.)\end{tabular}                                                           \\ 
\hline
CountLineBlank                                                                         & Number of blank lines                                                                                                                                         \\ 
\hline
Keywords                                                                               & \begin{tabular}[c]{@{}l@{}}The ratio between the number of \\ language keyword tokens to the \\ number of total tokens\end{tabular}                           \\ 
\hline
\begin{tabular}[c]{@{}l@{}}Operators in if, else, \\ and while statements\end{tabular} & \begin{tabular}[c]{@{}l@{}}The ratio between the number of \\ operators in if, else, and while sta-\\ -tements to the number of total \\ tokens\end{tabular}  \\
\hline
\end{tabular}
\end{adjustbox}
\end{table}


We used widely-used machine learning classifiers by researchers~\cite{Esteves2020Understanding, alikhashashneh2018using, medeiros2020vulnerable, li2023commit} such as Logistic Regression (LR)~\cite{lavalley2008logistic}, K-Nearest Neighbor (KNN)~\cite{peterson2009k}, Multi Layer Perceptron (MLP)~\cite{riedmiller2014multi}, Support Vector Machine (SVM)~\cite{riedmiller2014multi}, Random Forests (RF)~\cite{breiman2001random}, Decision Tree (DT)~\cite{suthaharan2016decision}, Gradient Boost (GB)~\cite{friedman2002stochastic}, and Extreme Gradient Boost (XGB)~\cite{chen2016xgboost}. We performed hyper-parameter tuning on each trained model to optimize the performance using random grid search~\cite{Jiang_2021}. 

\subsection{Machine Learning Classifiers with Code Embedding}
In this study, we also leveraged code embeddings to capture the information in source code with the goal of better distinguishing AI-generated and human-written code. We experimented with a pre-trained source code embedding model. Specifically, CodeT5+ 110M embedding model~\cite{wang2023codet5+} was selected because it was the latest code embedding model at the time we conducted the experiments. The generated embeddings were used as the features to train machine learning models. In order to generate the embeddings that capture different aspects of source code, we selected three code representations, namely, \textit{Code Only}, \textit{AST Only}~\cite{guo2022unixcoder}, and \textit{Combined} where we separated the concatenation of two representations with a special separator token following~\cite{guo2022unixcoder}. The goal was to explore the embeddings of various representations of source code (i.e., structural and textual information) to investigate the most distinguishing representations captured by embeddings. 
We used the same machine learning algorithms and steps used in Section \ref{sec:ml}, including random grid search. Figure \ref{fig:embcls} shows the overview of this approach. 

\begin{figure}[htp]
    \centering
    \includegraphics[width=0.7\columnwidth]{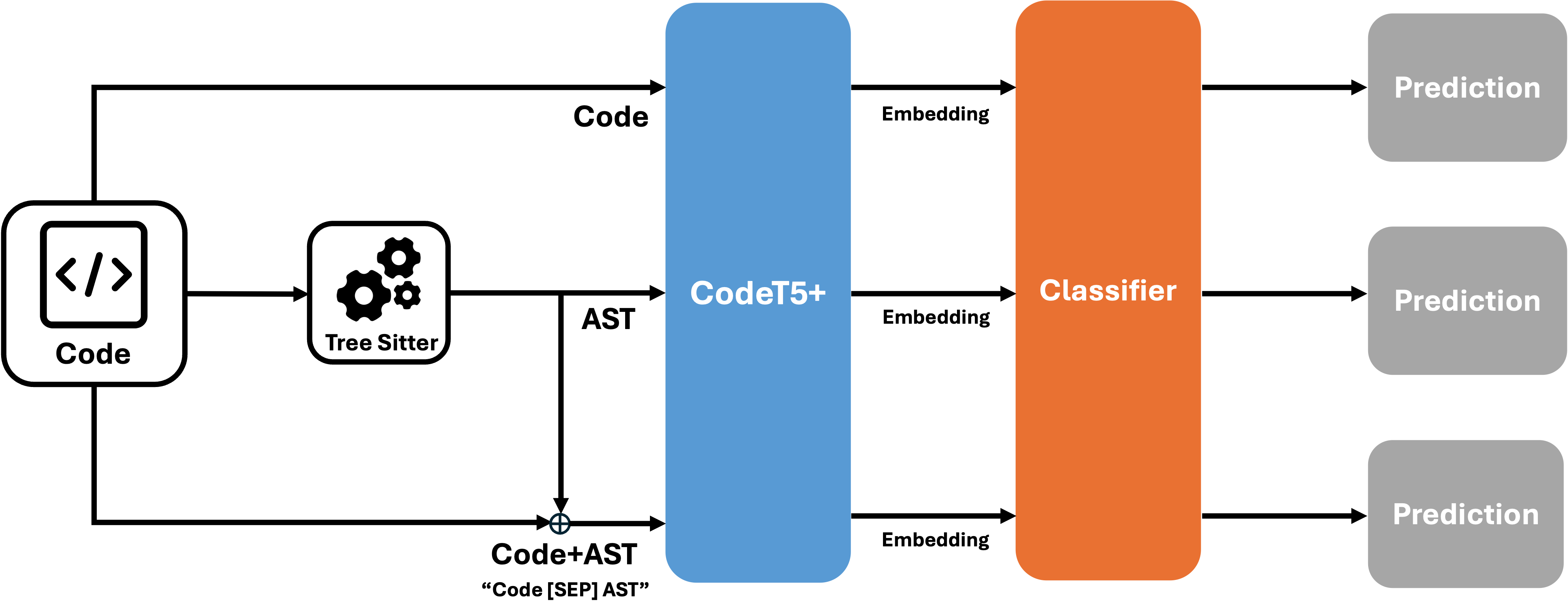}
    \caption{Overview of Machine Learning Classifiers with Embeddings}
    \label{fig:embcls}
\end{figure}

Furthermore, to explore performance inconsistencies across various approaches, we compared the similarity between AI-generated and human-written code to explain classification performance. We used semantic embeddings of code, which capture the underlying meaning of the code~\cite{wang2021codet5, allamanis2018learning, tufano2018deep}, providing a robust basis for comparison and understanding the differences between the two classes. We computed the cosine similarity between the code embeddings of AI-generated and human-written code from the same specification in our dataset and averaged these similarity values. The averaged cosine similarities between AI-generated and human-written code embeddings were then used to compare semantic similarity across the different LLMs. This process was carried out respectively for each of the four LLMs used in our study.

In addition, given that discrepancies between training and testing datasets can affect classification performance, we investigated these differences to understand why performance degrades in the \textit{``Across"} evaluation setting compared to the \textit{``Within"} evaluation setting. We focused on the cosine similarity of \textit{AST Only} embeddings, since models trained with these embeddings performed best in the \textit{``Within"} evaluation setting. In the \textit{``Within"} setting, we averaged the \textit{AST Only} embeddings separately for each dataset's training and testing splits. We then measured the cosine similarity between these averaged embeddings. We followed a similar process for the \textit{``Across"}  setting: averaging the \textit{AST Only} embeddings for the training and testing splits but comparing the cosine similarity with splits from different datasets. We compared 30 training-testing split combinations for each LLM. Finally, we averaged the cosine similarity values for each evaluation setting and compared the two sets of values.

\subsection{Ablation Study}
\label{sec:ablation}
Since our embedding-based machine learning models with \textit{AST Only} perform the best among all other approaches (Section ~\ref{sec:resembedding}), we conducted an ablation study to investigate the impact of different source code features on the models' performance. To do so, we first took all the 30 code features that are applicable to C++, Java and Python in Section \ref{sec:ml}. Among these features, we selected the ones that we can modify in the code without altering the code logic, namely \textit{Comment Lines}, \textit{Variable Names}, \textit{Method Names}, and \textit{Blank Lines}. We eliminated \textit{Blank Lines} from our experiments since removing blank lines does not have any impact on the AST representation of the source code.



Next, we established the following code variant types that do not affect the code logic, based on each of the three features:

\begin{itemize}
    \item Code with no comment line
    \item Code with uniform variable names (prefixed with 'var\_' and numbered sequentially, starting from var\_1)
    \item Code with uniform method names (prefixed with 'func\_' and numbered sequentially, starting from func\_1)
\end{itemize}

In order to create the mentioned variants, we used Tree-Sitter to parse the AST and change its relevant nodes. To create \textit{Code with no comment line}, we removed the \textit{comment} and \textit{block\_comment} nodes from the AST. To make the function names uniform (\textit{Code with uniform method names}), we changed the \textit{method\_declaration} or \textit{function\_definition} nodes given that each programming language has specific AST node types for function declaration/definition. Some language-specific functions such as the "main" method in C++ and Java, or the constructor and deconstructor methods (i.e. "\_\_init\_\_" in Python) remained untouched. For renaming the variables (\textit{Code with uniform variable names}), we performed a similar approach as we did for \textit{Code with uniform method names}. Figure \ref{fig:v34} shows an example of \textit{Code with uniform variable names}.

Different AST nodes were changed based on the AST structure and node types of each programming language. For instance, we changed the nodes, including "identifier", "pattern\_list", "assignment", "typed\_parameter" for Python code, while we modified "local\_variable\_declaration", "formal\_parameter" nodes for Java and "init\_declarator" for C++. The complete implementation is available in our replication package \cite{replicationpackage}. After generating the code variants, we produced the embeddings for the AST of these code variants. Finally, we trained machine learning classifiers with these \textit{AST Only} embeddings of the code variants for each variant type. Then we performed Welch's t-test~\cite{ruxton2006unequal} and calculated the effect size (Cohen's D~\cite{rosenthal1994parametric}) using the \textit{Average F1-scores} of each variant compared to those of the original code.



\begin{figure}[htp]
    \centering
    \begin{subfigure}[b]{0.2\textwidth}
         \centering
         \includegraphics[width=\textwidth]{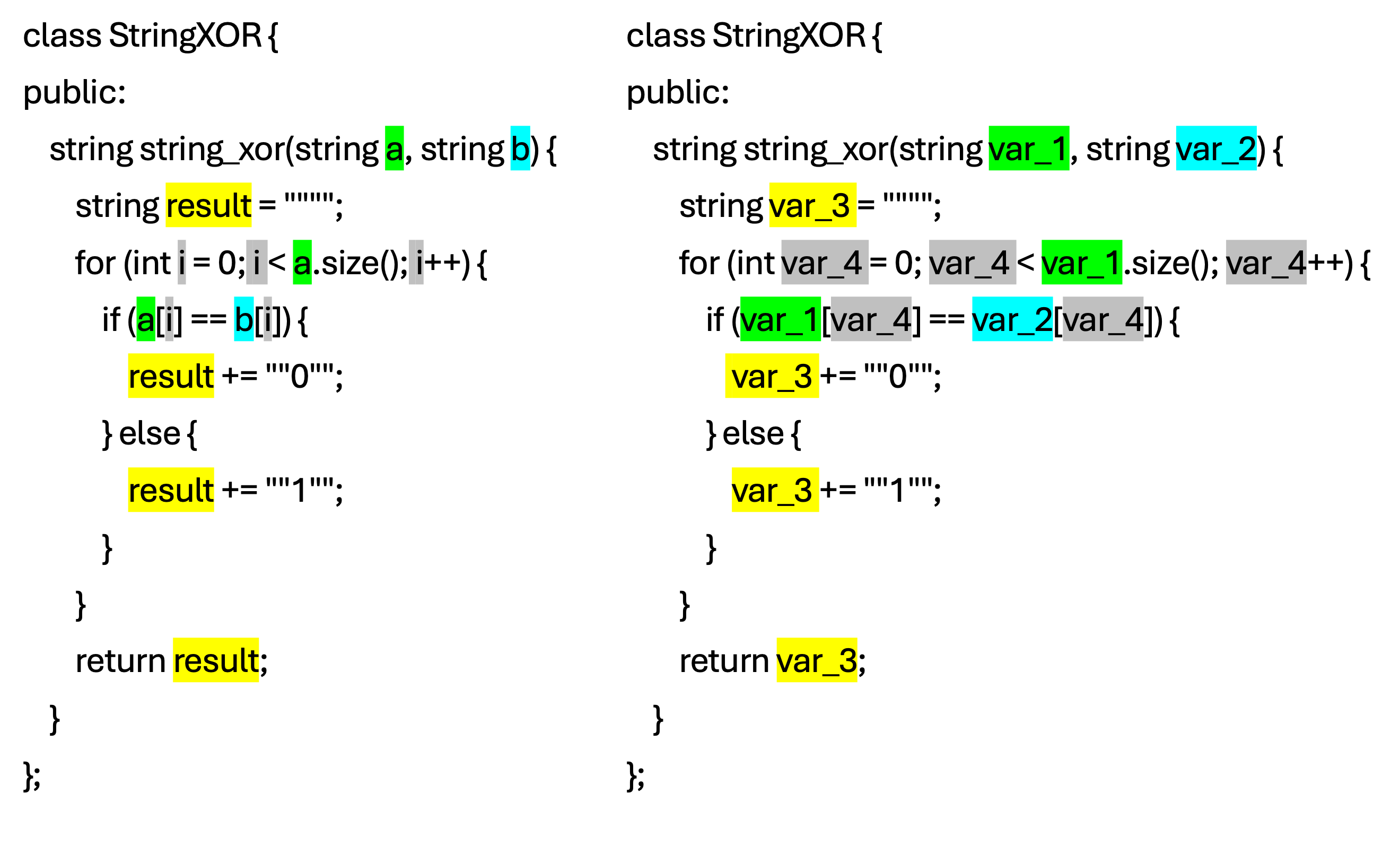}
         \label{fig:v3}
     \end{subfigure}
    \caption{Example of Code with uniform variable names Variant}
    \label{fig:v34}
\end{figure}

\section{Results}
\label{sec:results}


In this section, we organized the results of this study based on our research questions in Section \ref{sec:intro}. 
Due to space constraints, we present the results for the default temperature value. The result for 0 temperature is included on the companion website~\cite{replicationpackage}. We also include both the results for \textit{``Within"} and \textit{``Across"} evaluation settings to provide a comprehensive view of the results.

\begin{table*}
\centering
\caption{Performance of Existing AIGC Detectors (Default Temperature)}
\label{tab:aigc1}
\begin{adjustbox}{max width=\textwidth}
\scriptsize
\begin{tabular}{l|l|l|l|l|l|l|l|l|l|l|l|l|l|l|l|l|l|l|l|l} 
\hline
\multirow{2}{*}{}             & \multicolumn{4}{c|}{\textbf{ChatGPT}}                         & \multicolumn{4}{c|}{\textbf{Gemini Pro}}                      & \multicolumn{4}{c|}{\textbf{GPT-4}}                           & \multicolumn{4}{c|}{\textbf{Starcoder2-Instruct }}            & \multicolumn{4}{c}{\textbf{Mean of Every Datasets}}            \\ 
\cline{2-21}
                              & \textbf{ACC} & \textbf{TPR} & \textbf{TNR} & \textbf{AVG\_F1} & \textbf{ACC} & \textbf{TPR} & \textbf{TNR} & \textbf{AVG\_F1} & \textbf{ACC} & \textbf{TPR} & \textbf{TNR} & \textbf{AVG\_F1} & \textbf{ACC} & \textbf{TPR} & \textbf{TNR} & \textbf{AVG\_F1} & \textbf{ACC} & \textbf{TPR} & \textbf{TNR} & \textbf{AVG\_F1}  \\ 
\hline
\textbf{GPT2 Output Detector} & 45.61        & 0.00         & 100.00       & 31.88            & 36.66        & 0.00         & 100.00       & 34.54            & 46.65        & 49.07        & 53.07        & 32.25            & 50.22        & 26.49        & 73.95        & 45.54            & 48.86        & 18.89        & 81.75        & 36.05             \\ 
\hline
\textbf{DetectGPT}            & 46.44        & 2.95         & 100.00       & 33.61            & 38.97        & 7.21         & 99.51        & 39.13            & 43.71        & 43.31        & 52.26        & 37.24            & 52.64        & 7.26         & 98.01        & 39.70            & 48.99        & 15.18        & 87.45        & 37.42             \\ 
\hline
\textbf{Sapling}              & 55.17        & 94.46        & 11.60        & 44.10            & 33.79        & 92.73        & 7.83         & 37.25            & 45.90        & 43.62        & 35.09        & 35.55            & 52.18        & 47.49        & 56.88        & 50.81            & 49.97        & 69.57        & 27.85        & 41.93             \\ 
\hline
\textbf{GPTZero}              & 53.13        & 85.00        & 0.83         & 35.22            & 45.34        & 100.00       & 0.00         & 31.16            & 38.82        & 39.65        & 29.85        & 34.98            & 53.87        & 28.74        & 79.00        & 47.39            & 47.83        & 63.35        & 27.42        & 37.19             \\ 
\hline
\textbf{GLTR}                 & 44.53        & 48.73        & 44.64        & 41.63            & 41.83        & 45.32        & 68.46        & 47.80            & 51.83        & 45.48        & 46.59        & 46.59            & 47.21        & 47.08        & 47.34        & 44.22            & 50.83        & 46.65        & 51.76        & 45.06             \\
\hline
\end{tabular}
\end{adjustbox}
\end{table*}

\subsection{RQ1: How do existing AIGC detectors perform on detecting AI-generated source code?}
To answer RQ1, we evaluate the performance of existing AIGC detectors by running them on the testing splits of our datasets (Section \ref{sec:metrics}). As mentioned in Section \ref{sec:method}, five AIGC detectors for AI-generated natural language text (GPTZero, GPT-2 Output Detector, DetectGPT, GLTR, and Sapling) and a state-of-the-art AI-generated source code detector, GPTSniffer, are included as baseline approaches that we compare ours with in this study.

Table \ref{tab:aigc1} (AVG\_F1 stands for \textit{Average F1-score}) shows the performance of all five AIGC detectors for natural language text when the temperatures of the LLMs that generated the code are set as the default value (1 for ChatGPT, GPT-4, and Starcoder2-Instruct, 0.9 for Gemini Pro). We average the values of all the metrics across all datasets whose generation LLM is the same to obtain an overview of the performance. We find that the performance of detecting code when the generation LLMs' temperatures are set to 0 and the default value is similar, so we put the results when the temperatures of the LLMs are 0 to our replication package~\cite{replicationpackage} due to space constraints. Similar to what researchers have found~\cite{nguyen2023snippet,pan2024assessing}, their \textit{Accuracy} is mostly less than 0.6, indicating their ineffectiveness in detecting AI-generated source code. Since source code has unique linguistic syntax and writing styles that are different from natural language, the reasons for such low performance in detecting AI-generated source code could include that these AIGC detectors were trained with only natural language texts. 


Moreover, a certain AIGC detector can have a different performance in detecting code generated by different LLMs. For example, DetectGPT tends to classify human-written, ChatGPT-generated, and Gemini Pro-generated code as AI-generated based on the high \textit{TNR} and low \textit{TPR} values, but it does not have this issue when identifying GPT-4-generated code. We also find that some techniques tend to classify both human-written and AI-generated code as human-written code, such as GPTZero, on the datasets with code generated by Gemini Pro and ChatGPT. Code generated by Starcoder2-Instruct showed a relatively higher average F1-score compared to code generated by other LLMs across all AIGC detectors. Overall, all of them show limited effectiveness in detecting AI-generated code. 


In addition, the \textit{Average F1-score} across different generative LLMs does not show a significant difference, which indicates that AIGC detectors perform poorly regardless of the LLMs used to generate the AI-generated source code. We also find a similar trend in terms of the temperature settings of the generative LLMs~\cite{peng2023towards,geng2024large} (i.e,. LLMs used for code generation). Therefore, neither the generative LLMs nor their temperature settings affect the capability of existing AIGC detectors to detect AI-generated source code. 

\begin{mdframed}[roundcorner=10pt]
\textbf{Observation 1: Existing natural language AIGC detectors perform poorly in classifying human-written and AI-generated source code.} 
\end{mdframed}

As for GPTSniffer, a state-of-the-art AIGC detector fine-tuned with source code, we provide its performance results on all our datasets in Table \ref{tab:sniffer1}. While GPTSniffer has been fine-tuned on ChatGPT-generated Java code, it still shows poor performance in detecting source code from different datasets containing code written in Java (i.e. HumanEval-Java and CodeSearchNet-Java) other than its training data, even for Java code generated by ChatGPT. For example, it only shows the \textit{Accuracy} of 32.26 and \textit{Average F1-score} of 24.39 on Java code generated by ChatGPT in HumanEval dataset. When it comes to other programming languages, it shows limited performance on detecting Python code (i.e. \textit{Accuracy} of 44.85 and \textit{Average F1-score} of 30.96 on MBPP dataset with code generated by ChatGPT). Surprisingly, it shows a superior performance in detecting AI-generated code in C++ where it achieves \textit{Accuracy} of 100 and \textit{Average F1-score} of 100 on HumanEval-C++ dataset with ChatGPT-generated code. The reason could be that the testing split of this dataset only contains 33 data instances. A relatively smaller number of instances may have led to overfitting. 


For detecting code generated by LLMs other than ChatGPT, its \textit{Accuracy} fluctuates around 50. Moreover, it tends to classify code as AI-generated based on the high \textit{TNR} and low \textit{TPR} values. In addition, we also trained and tested our best-performing machine learning models (Section \ref{sec:resembedding}) on GPTSniffer's data, and we achieved similar performance (\textit{Average F1-score} of 1). Thus, we believe that GPTSniffer does not show a decent performance on datasets from a different domain (i.e., open-source projects in Github), code snippets written in programming languages other than Java, and generated by LLMs other than ChatGPT, which severely undermines its applicability to AI-generated code detection.

\begin{mdframed}[roundcorner=10pt]
\textbf{Observation 2: A state-of-the-art AIGC detector for source code, GPTSniffer, still performs poorly in classifying human-written and AI-generated source code.} 
\end{mdframed}




\begin{table}
\centering
\caption{The Evaluation Of GPTSniffer On Our Datasets (Default Temperature)}
\label{tab:sniffer1}
\begin{adjustbox}{max width=\textwidth}
\tiny
\begin{tabular}{l|l|l|l|l} 
\hline
\textbf{Dataset-Generation LLM-Language} & \textbf{ACC}   & \textbf{TPR}   & \textbf{TNR}   & \textbf{AVG\_F1}  \\ 
\hline
MBPP-ChatGPT-Python                      & 44.85          & 0.00           & 100.00         & 30.96             \\ 
\hline
MBPP-Gemini Pro-Python                   & 49.23          & 0.00           & 100.00         & 32.99             \\ 
\hline
MBPP-GPT-4-Python                        & 47.69          & 0.00           & 100.00         & 32.29             \\ 
\hline
MBPP-Starcoder2-Instruct-Python                   & 72.95          & 52.04          & 93.87          & 71.72             \\ 
\hline
HumanEval-ChatGPT-Python                 & 60.61          & 0.00           & 100.00         & 37.74             \\ 
\hline
HumanEval-ChatGPT-Java                   & 32.26          & 0.00           & 100.00         & 24.39             \\ 
\hline
HumanEval-ChatGPT-C++                    & 100.00         & 100.00         & 100.00         & 100.00            \\ 
\hline
HumanEval-Gemini Pro-Python              & 54.55          & 0.00           & 100.00         & 35.30             \\ 
\hline
HumanEval-Gemini Pro-Java                & 65.62          & 0.00           & 100.00         & 39.63             \\ 
\hline
HumanEval-Gemini Pro-C++                 & 60.61          & 85.71          & 42.11          & 60.02             \\ 
\hline
HumanEval-GPT-4-Python                   & 54.55          & 0.00           & 100.00         & 35.30             \\ 
\hline
HumanEval-GPT-4-Java                     & 57.58          & 0.00           & 100.00         & 36.54             \\ 
\hline
HumanEval-GPT-4-C++                      & 96.97          & 95.00          & 100.00         & 96.87             \\ 
\hline
HumanEval-Starcoder2-Instruct-Python              & 50.00          & 0.00           & 100.00         & 33.34             \\ 
\hline
HumanEval-Starcoder2-Instruct-Java                & 50.00          & 0.00           & 100.00         & 33.34             \\ 
\hline
HumanEval-Starcoder2-Instruct-C++                 & 85.29          & 100.00         & 70.58          & 84.96             \\ 
\hline
CodeSearchNet-ChatGPT-Python             & 50.00          & 0.00           & 100.00         & 33.34             \\ 
\hline
CodeSearchNet-ChatGPT-Java               & 47.50          & 0.00           & 100.00         & 32.21             \\ 
\hline
CodeSearchNet-Gemini Pro-Python          & 50.00          & 0.00           & 100.00         & 33.34             \\ 
\hline
CodeSearchNet-Gemini Pro-Java            & 47.50          & 0.00           & 95.00          & 32.21             \\ 
\hline
CodeSearchNet-GPT-4-Python               & 50.00          & 0.00           & 100.00         & 33.34             \\ 
\hline
CodeSearchNet-GPT-4-Java                 & 35.90          & 0.00           & 93.33          & 26.42             \\ 
\hline
CodeSearchNet-Starcoder2-Instruct-Python          & 57.50          & 30.00          & 85.00          & 54.02             \\ 
\hline
CodeSearchNet-Starcoder2-Instruct-Java            & 52.50          & 30.00          & 75.00          & 49.96             \\ 
\hline
\textbf{Average}                         & \textbf{57.24} & \textbf{20.53} & \textbf{93.95} & \textbf{45.01}    \\
\hline
\end{tabular}
\end{adjustbox}
\end{table}

\subsection{RQ2: How can we improve the performance of AI-generated source code detection? (LLM-based approaches)}
\label{sec:res_llm}

To improve the effectiveness of detecting AI-generated source code, we first experiment with a variety of LLM-based approaches. In this section, we evaluate the performance of ChatGPT after zero-shot learning, in-context learning and fine-tuning (\textit{fine-tuned ChatGPT}) where we prompted/trained the model with three different representations of source code (Section \ref{sec:LLMbased}), namely, \textit{Code Only}, \textit{AST Only}, and \textit{Combined}. 



We average the values of all the metrics across all datasets whose generation LLM is the same to obtain an overview of the performance as in Table \ref{tab:overall_result_tempdef} and \ref{tab:overall_result_tempdef_across} for \textit{``Within" evaluation setting} and \textit{``Across" evaluation setting}, respectively. Our detailed results for both evaluation settings are provided in our replication package~\cite{replicationpackage}. For \textit{fine-tuned ChatGPT}, their \textit{Accuracy} and \textit{Average F1-score} can reach more than 80 on some datasets such as datasets with ChatGPT and GPT-4-generated code in \textit{``Within" evaluation setting}, which suggests that \textit{fine-tuned ChatGPT} with our datasets can detect AI-generated code significantly better than existing AIGC detectors. We also find that \textit{fine-tuned ChatGPT} performs better in identifying AI-generated code produced by LLMs with a high temperature than that of a low temperature. 

However, the fine-tuned models only show around 40 in terms of \textit{Accuracy} and \textit{Average F1-score} in \textit{``Across" evaluation setting} where the models are evaluated on the testing splits of the different datasets (Section \ref{sec:metrics}). For example, \textit{fine-tuned ChatGPT} with MBPP dataset whose AI-generated Python code is by Gemini Pro shows only 50 in \textit{Accuracy} when being tested on CodeSearchNet dataset whose AI-generated code is written in Java (default temperature for the generative LLM). This indicates that \textit{fine-tuned ChatGPT} using \textit{Code Only} as the input still lack generalizability across code snippets from different domains and different programming languages.


Similar to \textit{fine-tuned ChatGPT} on \textit{Code Only}, \textit{fine-tuned ChatGPT} on \textit{AST Only} also significantly outperforms its zero-shot and in-context learning counterparts. However, the \textit{fine-tuned ChatGPT} performs significantly worse when using \textit{AST Only} as input than that of \textit{Code Only}. For example, the \textit{Average F1-score} for the \textit{fine-tuned ChatGPT} with human-written and ChatGPT-generated code using \textit{AST Only} is 58.96 
when the default temperature is used for code generation, while the score reaches 81.79 
for \textit{Code Only}. In addition, \textit{fine-tuned ChatGPT} on \textit{AST Only} shows poor performance in \textit{``Across" evaluation setting} (mean \textit{Average F1-score} of around 43 across all datasets when the generative LLMs' temperatures are 0, and 44 for default temperatures), suggesting limited generalizability. 

When we concatenate \textit{Code Only} and \textit{AST Only} as input (i.e., \textit{Combined}), we observe similar mean \textit{Average F1-scores} to those when \textit{Code Only} is used as input in zero-shot learning and in-context learning. For \textit{fine-tuned ChatGPT}, there is a notable performance drop compared to when \textit{Code Only} is used as input. For example, \textit{fine-tuned ChatGPT} on \textit{Code Only} outperform that of \textit{Combined} by more than 10 in terms of mean \textit{Average F1-scores} in the default temperature setting. Thus, we believe that AST representation of source code may not be suitable as the input for fine-tuning ChatGPT to detect AI-generated code. 


Interestingly, identifying source code generated by Gemini Pro presents the lowest mean \textit{Average F1-score}, suggesting it is particularly challenging to identify the code generated by this generative LLM using \textit{fine-tuned ChatGPT}. For example, \textit{fine-tuned ChatGPT} only has around 60 in \textit{Average F1-scores} and \textit{Accuracy} when detecting Gemini Pro-generated code, while it shows 70-80 when detecting ChatGPT, GPT-4, or Starcoder2-Instruct generated code when the temperature is set as 0.

\begin{mdframed}[roundcorner=10pt]
\textbf{Observation 3: \textit{fine-tuned ChatGPT} significantly outperforms zero shot and in-context learning. In addition, AST representation is not suitable as the input for \textit{fine-tuned ChatGPT} to detect AI-generated code.} 
\end{mdframed}





\subsection{RQ2: How can we improve the performance of AI-generated source code detection? (Machine Learning Classifiers with Static Code Metrics)}
\label{sec:mlstatic_res}

We evaluated the performance of the machine learning classifiers on the testing splits of the datasets. 
RF model shows the best performance when the temperature is set to 0, while GB shows the highest mean \textit{Average F1-score} in the default temperature setting. Similar to LLM-based approaches, we average the values of all the metrics across all datasets whose generation LLM is the same to obtain an overview of the performance. The results obtained under \textit{``Within" evaluation setting} are presented in Table \ref{tab:overall_result_tempdef} and the results under the \textit{``Across" evaluation setting} are presented in Table \ref{tab:overall_result_tempdef_across}.

The difference in \textit{Accuracy} and \textit{Average F1-score} across different generative LLMs shows that machine learning-based detection techniques have varied effectiveness in detecting code generated by different LLMs. For example, when the generative LLMs' temperatures are set to 0, the RF classifier has more than 80 in \textit{Average F1-score} and \textit{Accuracy} for detecting ChatGPT-generated code, while it only has around 66 in detecting Gemini Pro-generated code. However, these classifiers still outperform existing AIGC detectors. We also find that our machine learning classifiers tend to perform better (i.e., higher overall mean \textit{Average F1-score}) on detecting code generated with higher temperature (i.e., default temperature) than code generated with a temperature of 0. 

In \textit{``Across" evaluation setting}, the machine learning classifiers have a mean \textit{Average F1-score} of around 50 for both temperature of 0 and default temperature, which indicates the limited performance of the trained machine learning models when they are tested on different programming languages and datasets from other sources.







\begin{mdframed}[roundcorner=10pt]
\textbf{Observation 4: The machine learning classifiers trained with static code features can identify AI-generated code. However, they show varied effectiveness in detecting code generated by different LLMs.} 
\end{mdframed}

\subsection{RQ2: How can we improve the performance of AI-generated source code detection? (Machine Learning Classifiers with Embeddings)}
\label{sec:resembedding}
Instead of using static code features as in Section \ref{sec:mlstatic_res}, we adopted the code embeddings generated from \textit{Code Only}, \textit{AST Only}, and \textit{Combined} (Section \ref{sec:LLMbased}) using CodeT5+ as the feature vectors to train our machine learning models. We take SVM for the embedding of \textit{AST Only}, MLP for \textit{Code Only}, and LR for \textit{Combined} when the generative LLMs' temperatures are set to 0. In the default temperature setting, we select LR for all three representations. 



We highlight the highest scores across all approaches in Table \ref{tab:overall_result_tempdef}. Our findings are: machine learning models trained with the embeddings of \textit{AST Only} demonstrate the best performance across all the other approaches that we experimented with in this study, indicating that machine learning models are able to capture the difference between AI-generated and human-written code. The mean \textit{Average F1-score} is 81.44 when the temperature is set to 0, and 82.55 when the temperature is set to the default value. Thus, our machine learning models trained with embeddings significantly outperform AIGC detectors, including GPTSniffer.



However, these models still show inconsistent performance in identifying AI-generated code by different LLMs, similar to LLM-based and machine learning models trained with static code features. For example, their \textit{Accuracy} reaches nearly 90 (i.e., 89.89) when being trained and evaluated on human-written and ChatGPT-generated code, while it's only 73.36 for Gemini Pro, given the temperature is set as default.

Our results on exploring performance inconsistencies across various approaches show that the cosine similarity between AI-generated and human-written code embeddings were as follows: 73.87 for ChatGPT, 77.58 for Gemini Pro, 76.56 for GPT-4, and 75.20 for Starcoder2-Instruct. These values indicate a high degree of semantic similarity between the code embeddings of AI-generated and human-written code. Consistent with our result that code generated by Gemini Pro was the hardest to detect, the cosine similarity between embeddings of AI-generated and human-written code was the highest for Gemini Pro. This indicates that Gemini Pro's code is semantically closest to human-written code, making it more difficult to distinguish. In contrast, we found that the cosine similarity for code generated by ChatGPT and Starcoder2-Instruct was the lowest, suggesting greater semantic differences. This aligns with our finding that code generated by ChatGPT and Starcoder2-Instruct was the easiest to detect. Therefore, the varying semantic differences between AI-generated and human-written code across different LLMs may account for the performance inconsistencies observed in code detection. 

As for the model's generalizability, the results of the \textit{``Across" evaluation setting} show that it still struggles to detect AI-generated code written in other programming languages and from other domains (around 42 in terms of mean \textit{Average F1-score} across all datasets whose generative LLMs' temperatures are set to 0, and 45 for default temperature).
The results of analyzing the discrepancies of \textit{AST Only} embeddings between training and testing splits are shown in Table \ref{tab:within_vs_across_cossim}. On average, the cosine similarity in the \textit{``Across"} evaluation setting was about 20\% lower than in the \textit{``Within"} evaluation setting (77.87 vs. 98.68). This finding suggests that dissimilarities between training and testing datasets contribute to performance degradation in the \textit{``Across"} setting. However, further in-depth analysis is needed to fully understand all the factors affecting performance in the \textit{``Across"} setting.

\begin{mdframed}[roundcorner=10pt]
\textbf{Observation 5: The machine learning models trained with code embeddings of \textit{AST Only} show the best performance among all the approaches. However, they show varied effectiveness in detecting code generated by different LLMs.} 
\end{mdframed}






\begin{table*}[ht]
\centering
\caption{Overall Performance Comparison - Within (Default Temperature)}
\label{tab:overall_result_tempdef}
\begin{adjustbox}{max width=\textwidth}
\scriptsize
\begin{tabular}{l|l|l|l|l|l|l|l|l|l|l|l|l|l|l|l|l|l|l|l|l} 
\hline
\multirow{2}{*}{}             & \multicolumn{4}{c|}{\textbf{ChatGPT}}                                      & \multicolumn{4}{c|}{\textbf{Gemini Pro}}                                   & \multicolumn{4}{c|}{\textbf{GPT-4}}                                         & \multicolumn{4}{c|}{\textbf{Starcoder2-Instruct}}                                   & \multicolumn{4}{c}{\textbf{Overall Average}}                               \\ 
\cline{2-21}
                              & \textbf{ACC} & \textbf{TPR} & \textbf{TNR} & \textbf{AVG\_F1} & \textbf{ACC} & \textbf{TPR} & \textbf{TNR} & \textbf{AVG\_F1} & \textbf{ACC} & \textbf{TPR} & \textbf{TNR} & \textbf{AVG\_F1} & \textbf{ACC} & \textbf{TPR} & \textbf{TNR} & \textbf{AVG\_F1} & \textbf{ACC} & \textbf{TPR} & \textbf{TNR} & \textbf{AVG\_F1}  \\ 

\hline
\textbf{GPT-3.5 (Zero shot)-Code }     & 52.64 & \textbf{99.90} & 2.57  & 36.60 & 51.67 & \textbf{99.81} & 0.09  & 34.10 & 54.15 & \textbf{100.00} & 0.05  & 35.11 & 52.11 & \textbf{99.93} & 0.68  & 34.78 & 52.64 & \textbf{99.91} & 0.85  & 35.15  \\ 
\hline
\textbf{GPT-3.5 (In-context)-Code}     & 51.71 & 81.31  & 18.50 & 44.51 & 41.19 & 66.64  & 18.01 & 37.20 & 55.58 & 83.80  & 25.76 & 49.80 & 49.53 & 80.21  & 18.12 & 42.94 & 49.50 & 77.99  & 20.10 & 43.61  \\ 
\hline
\textbf{GPT-3.5 (Fine-tuned)-Code}     & 82.18 & 89.13  & 77.24 & 81.79 & 70.41 & 82.23  & 60.81 & 68.60 & \textbf{88.89} & 93.33  & \textbf{84.18} & \textbf{88.22} & 82.98 & 89.34  & 77.61 & 82.26 & 81.12 & 88.51  & 74.96 & 80.22  \\ 
\hline
\textbf{GPT-3.5 (Zero shot)-AST}       & 48.16 & 69.77  & 19.30 & 40.92 & 48.56 & 74.73  & 25.81 & 44.19 & 43.92 & 64.61  & 21.51 & 39.85 & 48.19 & 70.53  & 25.90 & 42.73 & 47.21 & 69.91  & 23.13 & 41.92  \\ 
\hline
\textbf{GPT-3.5 (In-context)-AST}      & 50.77 & 82.24  & 13.69 & 41.07 & 40.59 & 73.83  & 12.42 & 35.21 & 53.32 & 83.38  & 21.72 & 46.91 & 49.19 & 70.38  & 27.55 & 42.47 & 48.47 & 77.46  & 18.85 & 41.41  \\ 
\hline
\textbf{GPT-3.5 (Fine-tuned)-AST}      & 64.18 & 46.71  & 81.82 & 58.96 & 55.86 & 56.76  & 59.55 & 53.07 & 65.64 & 83.22  & 48.96 & 62.61 & 60.10 & 53.50  & 68.20 & 54.53 & 61.44 & 60.05  & 64.63 & 57.29  \\ 
\hline
\textbf{GPT-3.5 (Zero shot)-Code+AST}  & 55.50 & 97.78  & 4.29  & 38.78 & 44.65 & 97.54  & 0.69  & 31.30 & 52.05 & 98.19  & 1.77  & 35.51 & 52.56 & 97.99  & 2.28  & 35.89 & 51.19 & 97.87  & 2.26  & 35.37  \\ 
\hline
\textbf{GPT-3.5 (In-context)-Code+AST} & 46.27 & 50.71  & 40.43 & 42.90 & 37.90 & 52.01  & 22.01 & 32.70 & 52.57 & 61.81  & 43.50 & 50.25 & 49.10 & 59.15  & 38.30 & 45.66 & 46.46 & 55.92  & 36.06 & 42.88  \\ 
\hline
\textbf{GPT-3.5 (Fine-tuned)-Code+AST} & 75.25 & 91.68  & 57.12 & 72.57 & 63.06 & 85.23  & 44.26 & 59.37 & 75.37 & 90.42  & 61.87 & 71.72 & 67.09 & 73.66  & 61.43 & 61.79 & 70.19 & 85.25  & 56.17 & 66.36  \\ 
\hline
\textbf{Code Metrics (GB)}             & 78.72 & 84.81  & 73.62 & 78.28 & 72.78 & 81.10  & \textbf{65.89} & 72.41 & 76.03 & 78.82  & 72.52 & 75.57 & 74.01 & 80.86  & 67.02 & 73.67 & 75.84 & 81.58  & 70.68 & 74.98  \\ 
\hline
\textbf{Code Embedding-Code}           & 81.15 & 83.37  & 80.41 & 80.91 & 68.12 & 78.50  & 60.27 & 67.87 & 81.76 & 84.20  & 79.90 & 81.47 & 86.09 & 86.21  & \textbf{86.01} & 86.00 & 79.28 & 83.07  & 76.65 & 79.06  \\
\hline
\textbf{Code Embedding-AST}            & \textbf{89.89} & 91.76  & \textbf{88.34} & \textbf{89.80} & \textbf{73.36} & 85.30  & 64.02 & \textbf{72.97} & 82.26 & 83.49  & 81.12 & 82.10 & 85.40 & 85.79  & 84.96 & 85.33 & \textbf{82.72} & 86.59  & \textbf{79.61} & \textbf{82.55}  \\ 
\hline
\textbf{Code Embedding-Code+AST}      & 84.85 & 84.67  & 85.45 & 84.46 & 70.05 & 79.90  & 62.28 & 69.81 & 82.16 & 81.88  & 82.25 & 82.00 & \textbf{87.12} & 90.28  & 83.93 & \textbf{87.10} & 81.05 & 84.18  & 78.48 & 80.84  \\
\hline
\end{tabular}
\end{adjustbox}
\end{table*}

\begin{table*}
\centering
\caption{Overall Performance Comparison - Across (Default Temperature)}
\label{tab:overall_result_tempdef_across}
\begin{adjustbox}{max width=\textwidth}
\scriptsize
\begin{tabular}{l|l|l|l|l|l|l|l|l|l|l|l|l|l|l|l|l|l|l|l|l} 
\hline
\multirow{2}{*}{}                      & \multicolumn{4}{c|}{\textbf{ChatGPT }}                        & \multicolumn{4}{c|}{\textbf{Gemini Pro }}                     & \multicolumn{4}{c|}{\textbf{GPT-4 }}                          & \multicolumn{4}{c|}{\textbf{Starcoder2-Instruct}}                     & \multicolumn{4}{c}{\textbf{Overall Average}}                   \\ 
\cline{2-21}
                                       & \textbf{ACC} & \textbf{TPR} & \textbf{TNR} & \textbf{AVG\_F1} & \textbf{ACC} & \textbf{TPR} & \textbf{TNR} & \textbf{AVG\_F1} & \textbf{ACC} & \textbf{TPR} & \textbf{TNR} & \textbf{AVG\_F1} & \textbf{ACC} & \textbf{TPR} & \textbf{TNR} & \textbf{AVG\_F1} & \textbf{ACC} & \textbf{TPR} & \textbf{TNR} & \textbf{AVG\_F1}  \\ 
\hline
\textbf{GPT-3.5 (In-context)-Code}     & 51.76        & 66.13        & 33.17        & 44.29            & 47.38        & 81.75        & 19.50        & 40.92            & 55.12        & 86.78        & 19.75        & 46.24            & 48.36        & 82.17        & 14.55        & 40.01            & 50.66        & 79.21        & 21.74        & 42.87             \\ 
\hline
\textbf{GPT-3.5 (Fine-tuned)-Code}     & 52.11        & 56.50        & 45.63        & 44.80            & 50.64        & 72.27        & 34.30        & 43.44            & 53.26        & 49.10        & 54.55        & 46.01            & 51.26        & 12.98        & 89.53        & 37.61            & 51.82        & 47.71        & 56.00        & 42.96             \\ 
\hline
\textbf{GPT-3.5 (In-context)-AST}      & 46.69        & 40.21        & 53.53        & 43.93            & 51.05        & 49.21        & 51.95        & 48.47            & 51.38        & 42.30        & 59.02        & 48.74            & 47.95        & 40.99        & 54.95        & 45.78            & 49.27        & 43.18        & 54.86        & 46.73             \\ 
\hline
\textbf{GPT-3.5 (Fine-tuned)-AST}      & 52.77        & 39.29        & 64.67        & 44.33            & 51.17        & 47.87        & 53.49        & 46.27            & 51.47        & 60.54        & 40.53        & 44.67            & 52.93        & 30.49        & 75.51        & 39.75            & 52.09        & 44.55        & 58.55        & 43.76             \\ 
\hline
\textbf{GPT-3.5 (In-context)-Code+AST} & 47.50        & 54.70        & 41.11        & 44.16            & 48.08        & 60.96        & 35.71        & 45.02            & 54.08        & 60.78        & 44.40        & 50.46            & 49.95   \textcolor{blue} {}    & 59.66        & 40.10        & 47.82            & 49.90        & 59.02        & 40.33        & 46.86             \\ 
\hline
\textbf{GPT-3.5 (Fine-tuned)-Code+AST} & 51.72        & 51.55        & 49.57        & 46.32            & 52.64        & 71.98        & 34.70        & 45.42            & 54.48        & 56.08        & 51.90        & 48.72            & 51.54        & 29.68        & 73.50        & 45.47            & 52.59        & 52.32        & 52.42        & 46.48             \\ 
\hline
\textbf{Code Metrics (GB)}             & 56.79        & 54.43        & 59.11        & 51.82            & 58.35        & 59.93        & 56.15        & 54.62            & 55.32        & 47.68        & 64.02        & 50.96            & 51.88        & 37.63        & 66.33        & 46.72            & 55.59        & 49.92        & 61.40        & 51.03             \\ 
\hline
\textbf{Code Embedding-Code}           & 51.15        & 39.01        & 63.72        & 42.20            & 57.23        & 50.91        & 62.70        & 50.72            & 52.19        & 35.71        & 69.11        & 43.38            & 53.47        & 38.67        & 69.42        & 46.83            & 53.51        & 41.08        & 66.24        & 45.78             \\ 
\hline
\textbf{Code Embedding-AST}            & 53.47        & 38.67        & 69.42        & 46.83            & 53.55        & 38.06        & 67.38        & 45.41            & 50.84        & 26.09        & 78.72        & 42.83            & 50.07        & 18.26        & 82.12        & 37.78            & 51.98        & 30.27        & 74.41        & 43.21             \\ 
\hline
\textbf{Code Embedding-Code+AST}       & 54.85        & 42.37        & 67.89        & 46.64            & 55.27        & 45.83        & 64.11        & 48.42            & 55.94        & 37.82        & 74.49        & 47.58            & 49.09        & 17.31        & 81.06        & 38.36            & 53.79        & 35.83        & 71.89        & 45.25             \\
\hline
\end{tabular}
\end{adjustbox}
\end{table*}

\begin{table*}
\centering
\caption{Comparison on the Characteristics of Training and Testing dataset - Within vs. Across (Default Temperature)}
\label{tab:within_vs_across_cossim}
\begin{adjustbox}{max width=\textwidth}
\begin{tabular}{l|l|l|l|l|l|l|l|l|l|l} 
\hline
\multirow{2}{*}{}                      & \multicolumn{2}{c|}{\textbf{ChatGPT }} & \multicolumn{2}{c|}{\textbf{Gemini Pro }} & \multicolumn{2}{c|}{\textbf{GPT-4 }} & \multicolumn{2}{c|}{\textbf{Starcoder2-Instruct}} & \multicolumn{2}{c}{\textbf{Overall Average }}  \\ 
\cline{2-11}
                                       & \textbf{Within} & \textbf{Across}      & \textbf{Within} & \textbf{Across}         & \textbf{Within} & \textbf{Across}    & \textbf{Within} & \textbf{Across}         & \textbf{Within} & \textbf{Across}              \\ 
\hline
\textbf{Cosine Similarity of AST Embeddings} & 98.61           & 78.76                & 98.65           & 75.18                   & 98.63           & 78.09              & 98.85           & 79.46                 & 98.68          & 77.87                     \\
\hline
\end{tabular}
\end{adjustbox}
\end{table*}

\subsection{RQ3: How do the source code features captured by embeddings contribute to the overall effectiveness? (Ablation Study)}
\label{sec:ablation_res}
In our ablation study, we generate the code variants as described in Section \ref{sec:ablation} and produce embeddings for these variants using CodeT5+. Specifically, we used the best-performing model across all the approaches we experimented with (the model with the highest mean \textit{Average F1-score}) in this ablation study as the AI-generated code detection model, namely, the GB classifier trained with the embedding of \textit{AST Only} with code generated by LLMs in default temperature setting. We use \textit{Best model} to represent this model for simplicity. Then, we compare this model's performance with the performance of the GB classifiers trained on the embeddings of the code variants of the three variant types. The goal is to investigate whether the change of the code features captured by code embeddings can have an impact on the overall effectiveness of the \textit{Best model}.


The results in Table \ref{tab:ablation_result} show that the variants of \textit{Code with no comment line} have the most impact on the performance of the \textit{Best model} where it decreases the mean \textit{Average F1-score} by 3.82, while variants of \textit{Code with uniform variable names} and \textit{Code with uniform method names} have negligible effect. We conducted t-test and measured the effect size on the \textit{Average F1-score} of each variants against the \textit{Average F1-score} of the \textit{Best model} to see if there were any statistically significant differences between the values. For variants of \textit{Code with no comment line}, we observed p-value of 0.4543 
and effect size of 0.2178 
. The results show that the impact is statistically insignificant with a small effect size. 

\begin{mdframed}[roundcorner=10pt]
\textbf{Observation 6: Removing code comments has an impact on the effectiveness of our best model. However, the impact is statistically insignificant with a small effect size.} 
\end{mdframed}

\begin{table*}
\caption{The Result of Ablation Study}
\label{tab:ablation_result}
\centering
\begin{adjustbox}{max width=\textwidth}
\scriptsize
\begin{tabular}{l|l|l|l|l|l|l|l|l|l|l|l|l|l|l|l|l|l|l|l|l|l|l|l|l} 
\hline
\multirow{2}{*}{}       & \multicolumn{4}{c|}{\textbf{ChatGPT}}       & \multicolumn{4}{c|}{\textbf{Gemini Pro}}    & \multicolumn{4}{c|}{\textbf{GPT-4}}         & \multicolumn{4}{c|}{\textbf{Starcoder2-Instruct}} & \multicolumn{4}{c|}{\textbf{Overall Average}} & \multicolumn{4}{l}{\textbf{Difference (Compared to Base)}}  \\ 
\cline{2-25}\textcolor{blue}
{}                        & \textbf{ACC} & \textbf{TPR} & \textbf{TNR} & \textbf{AVG\_F1} & \textbf{ACC} & \textbf{TPR} & \textbf{TNR} & \textbf{AVG\_F1} & \textbf{ACC} & \textbf{TPR} & \textbf{TNR} & \textbf{AVG\_F1} & \textbf{ACC} & \textbf{TPR} & \textbf{TNR} & \textbf{AVG\_F1} & \textbf{ACC} & \textbf{TPR} & \textbf{TNR} & \textbf{AVG\_F1} & \textbf{ACC} & \textbf{TPR} & \textbf{TNR} & \textbf{AVG\_F1} \\ 
\hline
\textbf{Base AST Code Embedding} & 89.88 & 91.76 & 88.33 & 89.80   & 73.35 & 85.30 & 64.02 & 72.97   & 82.25 & 83.48 & 81.11 & 82.10   & 85.40 & 85.79 & 84.96 & 85.33   & 82.72 & 86.59 & 79.61 & 82.55        & -     & -     & -     & -                          \\ 
\hline
\textbf{Uniform Method Names}    & 88.36  & 92.89  & 85.07  & 88.33   & 75.30  & 88.93  & 64.78  & 74.97   & 80.90  & 82.56  & 79.59  & 80.83   & 87.22 & 90.05 & 84.33 & 87.20   & 82.95 & 88.61 & 78.44 & 82.83        & 0.22  & 2.02  & -1.17 & 0.29                       \\ 
\hline
\textbf{Uniform Variable Names}  & 85.49  & 87.80  & 83.49  & 85.40   & 75.07  & 85.35  & 67.06  & 74.56   & 83.52  & 86.66  & 80.51  & 83.39   & 85.39 & 88.73 & 81.95 & 85.34   & 82.37 & 87.14 & 78.25 & 82.17        & -0.36 & 0.55  & -1.36 & -0.38                      \\ 
\hline
\textbf{No Comment Line}         & 83.98  & 86.21  & 81.58  & 83.43   & 73.29  & 83.79  & 64.56  & 72.68   & 78.03  & 79.99  & 75.38  & 77.37   & 81.49 & 83.95 & 78.97 & 81.45   & 79.20 & 83.49 & 75.12 & 78.73        & -3.53 & -3.10 & -4.49 & -3.82                      \\
\hline
\end{tabular}
\end{adjustbox}
\end{table*}

\section{Discussion}
\label{sec:discussion}


Despite the fact that the LLMs such as ChatGPT have shown state-of-the-art performance using zero-shot and in-context learning in a variety of SE tasks, including code summarization~\cite{geng2024large}, code refinement~\cite{guo2024exploring}, and Android bug replay~\cite{feng2024prompting}, they both have relatively poor performance by showing around 40 in \textit{Average F1-scores} and \textit{Accuracy} across all datasets in detecting AI-generated code. Thus, we believe that even powerful LLMs such as ChatGPT lack the capability to effectively identify AI-generated source code if they are not fine-tuned, even when retrieved demonstration examples are provided in the prompt.

Based on our results in \ref{sec:results}, we are able to reach over 80 in terms of \textit{Accuracy} and \textit{Average F1-scores}. Compared to existing AIGC detectors and GPTSniffer, our models have achieved significantly better performance. However, there is still room for improvement since all our models perform poorly in the \textit{``Across" evaluation setting} where they fail to be effectively generalized to code written in programming languages other than the language of their training data and from different domains or sources. This is not surprising as it also happens to other SE tasks. For example, while there has been research improving the performance of defect prediction models for decades, the task can still be challenging when the models are evaluated on data from software projects other than the project that they are trained with~\cite{hosseini2017systematic}. Thus, further research from the community is required to devise generalizable detection techniques.

For using different representations of source code to train our models (\textit{Code Only}, \textit{AST Only}, and \textit{Combined}), there is no representation that always outperforms another across all our approaches. \textit{Code Only} shows the best performance in fine-tuning ChatGPT, while machine learning models trained with the embeddings of \textit{AST Only} achieve the highest \textit{Accuracy} and \textit{Average F1-scores} among all other approaches. This may suggest the need for a multi-modal representation of code for better AI-generated code detection, and further research is needed to explain this observation conclusively.


Our models have different performances, and they, in general, perform better in detecting code generated by ChatGPT, GPT-4, and Starcoder2-Instruct than that of Gemini Pro. Some probable reasons behind this could be that Gemini Pro tends to generate code that resembles what humans would write, or our features and embeddings could not capture the difference between Gemini Pro-generated code and human-written code. As LLMs are being built and trained at an ever-increasing rate with better performance, more effective approaches with distinguishing features or embeddings should be implemented to detect AI-generated code that's getting ever closer to human-written one.

Finally, in our ablation study (Section \ref{sec:ablation_res}), while removing comments from the source code, has the most impact on the \textit{Best model}'s performance, the impact is statistically insignificant.

\section{Threats to Validity}
\label{sec:ttv}

We have taken all reasonable steps to mitigate potential threats that could hamper the validity of this study. 

\textbf{Construct validity} It is possible that the prompt we used to generate the source code with the LLMs might have impacted the generated code quality. To mitigate this threat, we followed the best practices of prompt design and the setting of in-context examples~\cite{gao2023makes,ibm,awesomeprompt}, such as explicitly defining the persona of the LLMs, instructing the task definition, and providing necessary context information. However, we did not use any advanced prompting techniques, such as Chain-of-Thought, which could potentially improve the code quality. This omission may affect the validity of our results, as these techniques might yield better quality code and influence the detection outcomes. 

Another concern is the potential bias arising from the specifications and code in the datasets, which are collected from public platforms such as LeetCode and Github. ChatGPT, Gemini Pro, or Starcoder2-Instruct might have been trained on these datasets. 
Since LLMs are probabilistic generative models that create code based on specifications rather than retrieving it from their training datasets, the risk of bias in AI-generated code is likely reduced. However, this issue is more pronounced with human-written code, which is part of the dataset and thus more accessible for LLMs during their learning. LLMs may therefore detect human-written code more effectively than AI-generated code because they may have encountered human-written code during their pre-training phase. This potential bias could influence our findings.

\textbf{Internal validity} The generation of the LLMs, such as ChatGPT or Gemini Pro, is non-deterministic, especially when the temperature is set to 1. This may negatively impact the reproducibility of our study.

There might have been possible mistakes in implementing the baseline AIGC detectors that we compared with. To alleviate this threat, we directly used the code implementations published by the authors~\cite{pan2024assessing,nguyen2023snippet}. For commercial tools (i.e., GPTZero and Sapling), we used their publicly available API. 


\textbf{External validity} The findings in this study may only be valid for the three datasets and the three LLMs we selected. However, the code snippets and specifications in the datasets come from a diverse range of sources. The three LLMs are among the most widely-used LLMs by software developers for code generation and they are built by different organizations (i.e. OpenAI VS Google). Moreover, multiple programming languages that are widely adopted by software practitioners are considered in this study. Thus, we believe the concern about the generalizability of our findings is mitigated.
\section{Conclusion}
\label{sec:conclusion}

Our study identified that existing AIGC detectors perform poorly in detecting AI-generated source code. Our results indicated that current techniques for detecting AI-generated source code need to be improved, underscoring the need for further research in this area. From these findings, we suggested the following three different approaches for AI-generated source code detection: (1) LLM-based approach, (2) Machine Learning with Code Metrics, and (3) Machine Learning with Code Embeddings. We evaluated our approaches on multiple datasets from different sources, LLM, and languages. Among the approaches, Machine Learning with Code Embeddings yielded the highest mean \textit{Average F1-score} of 82.55 in \textit{"Within" evaluation setting}. We conducted an ablation study with our best-performing model to investigate the impact of different source code features on the model's performance. Our paper's findings unveil the current status of AI-generated source code detection and propose pathways for improvement.  


\bibliographystyle{IEEEtranS}
\bibliography{icse}

\begin{thebibliography}{10}
\providecommand{\url}[1]{#1}
\csname url@samestyle\endcsname
\providecommand{\newblock}{\relax}
\providecommand{\bibinfo}[2]{#2}
\providecommand{\BIBentrySTDinterwordspacing}{\spaceskip=0pt\relax}
\providecommand{\BIBentryALTinterwordstretchfactor}{4}
\providecommand{\BIBentryALTinterwordspacing}{\spaceskip=\fontdimen2\font plus
\BIBentryALTinterwordstretchfactor\fontdimen3\font minus \fontdimen4\font\relax}
\providecommand{\BIBforeignlanguage}[2]{{%
\expandafter\ifx\csname l@#1\endcsname\relax
\typeout{** WARNING: IEEEtranS.bst: No hyphenation pattern has been}%
\typeout{** loaded for the language `#1'. Using the pattern for}%
\typeout{** the default language instead.}%
\else
\language=\csname l@#1\endcsname
\fi
#2}}
\providecommand{\BIBdecl}{\relax}
\BIBdecl

\bibitem{chatgpt}
``Chatgpt: Optimizing language models for dialogue,'' https://openai.com/blog/chatgpt, 2022, 2022.

\bibitem{ibm}
``Ibm global ai adoption index 2022,'' https://www.ibm.com/watson/resources/ai-adoption, 2022, 2022.

\bibitem{awesomeprompt}
``awesome-chatgpt-prompts,'' https://github.com/f/awesome-chatgpt-prompts, 2023, 2023.

\bibitem{gpt-4}
``Gpt-4,'' https://openai.com/research/gpt-4, 2023, 2023.

\bibitem{gpt2detector}
``Gpt2detector,'' https://github.com/MacroChip/gpt-2-output-dataset, 2023, 2023.

\bibitem{gptzero}
``Gptzero,'' https://gptzero.me/, 2023, 2023.

\bibitem{sapling}
``Sapling,'' https://sapling.ai/ai-content-detector, 2023, 2023.

\bibitem{topprogram}
``The top programming languages 2023,'' https://spectrum.ieee.org/the-top-programming-languages-2023, 2023, 2023.

\bibitem{copilot}
``Github copilot,'' https://github.com/features/copilot, 2024, 2024.

\bibitem{gemini}
``Google gemini,'' https://gemini.google.com/, 2024, 2024.

\bibitem{openai}
``Pricing,'' https://openai.com/pricing, 2024, 2024.

\bibitem{replicationpackage}
``Replication package,'' 2024, \url{https://anonymous.4open.science/r/AI-Detector-3534/}.

\bibitem{und}
``Scitools understand,'' https://scitools.com/, 2024, 2024.

\bibitem{TIOBE}
``Tiobe index for march 2024,'' https://www.tiobe.com/tiobe-index/, 2024, 2024.

\bibitem{treesitter}
``Tree-sitter,'' https://tree-sitter.github.io/tree-sitter/, 2024, 2024.

\bibitem{akinwande2015variance}
O.~Akinwande, H.~Dikko, and S.~Agboola, ``Variance inflation factor: As a condition for the inclusion of suppressor variable(s) in regression analysis,'' \emph{Open Journal of Statistics}, vol.~05, pp. 754--767, 01 2015.

\bibitem{alikhashashneh2018using}
E.~A. Alikhashashneh, R.~R. Raje, and J.~H. Hill, ``Using machine learning techniques to classify and predict static code analysis tool warnings,'' in \emph{2018 IEEE/ACS 15th International Conference on Computer Systems and Applications (AICCSA)}, 2018, pp. 1--8.

\bibitem{aljehane2021determining}
S.~Aljehane, B.~Sharif, and J.~Maletic, ``Determining differences in reading behavior between experts and novices by investigating eye movement on source code constructs during a bug fixing task,'' in \emph{ACM Symposium on Eye Tracking Research and Applications}, 2021, pp. 1--6.

\bibitem{allamanis2018learning}
\BIBentryALTinterwordspacing
M.~Allamanis, M.~Brockschmidt, and M.~Khademi, ``Learning to represent programs with graphs,'' in \emph{International Conference on Learning Representations}, 2018. [Online]. Available: \url{https://openreview.net/forum?id=BJOFETxR-}
\BIBentrySTDinterwordspacing

\bibitem{asare2023github}
O.~Asare, M.~Nagappan, and N.~Asokan, ``Is github’s copilot as bad as humans at introducing vulnerabilities in code?'' \emph{Empirical Software Engineering}, vol.~28, no.~6, p. 129, 2023.

\bibitem{austin2021program}
J.~Austin, A.~Odena, M.~Nye, M.~Bosma, H.~Michalewski, D.~Dohan, E.~Jiang, C.~Cai, M.~Terry, Q.~Le \emph{et~al.}, ``Program synthesis with large language models,'' \emph{arXiv preprint arXiv:2108.07732}, 2021.

\bibitem{breiman2001random}
L.~Breiman, ``Random forests,'' \emph{Machine learning}, vol.~45, pp. 5--32, 2001.

\bibitem{brown2020language}
T.~Brown, B.~Mann, N.~Ryder, M.~Subbiah, J.~D. Kaplan, P.~Dhariwal, A.~Neelakantan, P.~Shyam, G.~Sastry, A.~Askell \emph{et~al.}, ``Language models are few-shot learners,'' \emph{Advances in neural information processing systems}, vol.~33, pp. 1877--1901, 2020.

\bibitem{buch2019learning}
L.~B{\"u}ch and A.~Andrzejak, ``Learning-based recursive aggregation of abstract syntax trees for code clone detection,'' in \emph{2019 IEEE 26th International Conference on Software Analysis, Evolution and Reengineering (SANER)}.\hskip 1em plus 0.5em minus 0.4em\relax IEEE, 2019, pp. 95--104.

\bibitem{chen2021evaluating}
M.~Chen, J.~Tworek, H.~Jun, Q.~Yuan, H.~P. d.~O. Pinto, J.~Kaplan, H.~Edwards, Y.~Burda, N.~Joseph, G.~Brockman \emph{et~al.}, ``Evaluating large language models trained on code,'' \emph{arXiv preprint arXiv:2107.03374}, 2021.

\bibitem{chen2016xgboost}
T.~Chen and C.~Guestrin, ``Xgboost: A scalable tree boosting system,'' in \emph{Proceedings of the 22nd acm sigkdd international conference on knowledge discovery and data mining}, 2016, pp. 785--794.

\bibitem{chen2018remarkable}
Z.~Chen and M.~Monperrus, ``The remarkable role of similarity in redundancy-based program repair,'' \emph{arXiv preprint arXiv:1811.05703}, 2018.

\bibitem{clemente2018predicting}
C.~Clemente, F.~Jaafar, and Y.~Malik, ``Is predicting software security bugs using deep learning better than the traditional machine learning algorithms?'' 07 2018, pp. 95--102.

\bibitem{nicad}
J.~R. Cordy and C.~K. Roy, ``The nicad clone detector,'' in \emph{2011 IEEE 19th International Conference on Program Comprehension}, 2011, pp. 219--220.

\bibitem{dai2015semi}
A.~M. Dai and Q.~V. Le, ``Semi-supervised sequence learning,'' \emph{Advances in neural information processing systems}, vol.~28, 2015.

\bibitem{dehaerne2022code}
E.~Dehaerne, B.~Dey, S.~Halder, S.~De~Gendt, and W.~Meert, ``Code generation using machine learning: A systematic review,'' \emph{Ieee Access}, vol.~10, pp. 82\,434--82\,455, 2022.

\bibitem{Dias2020Software}
\BIBentryALTinterwordspacing
E.~Dias~Canedo and B.~Cordeiro~Mendes, ``\BIBforeignlanguage{en}{Software requirements classification using machine learning algorithms},'' \emph{\BIBforeignlanguage{en}{Entropy}}, vol.~22, no.~9, p. 1057, 9 2020. [Online]. Available: \url{http://dx.doi.org/10.3390/e22091057}
\BIBentrySTDinterwordspacing

\bibitem{ding2022can}
\BIBentryALTinterwordspacing
Z.~Ding, H.~Li, W.~Shang, and T.-H.~P. Chen, ``Can pre-trained code embeddings improve model performance? revisiting the use of code embeddings in software engineering tasks,'' \emph{Empirical Softw. Engg.}, vol.~27, no.~3, may 2022. [Online]. Available: \url{https://doi.org/10.1007/s10664-022-10118-5}
\BIBentrySTDinterwordspacing

\bibitem{Esteves2020Understanding}
\BIBentryALTinterwordspacing
G.~Esteves, E.~Figueiredo, A.~Veloso, M.~Viggiato, and N.~Ziviani, ``\BIBforeignlanguage{en}{Understanding machine learning software defect predictions},'' \emph{\BIBforeignlanguage{en}{Automated Software Engineering}}, vol.~27, no. 3-4, pp. 369--392, 10 2020. [Online]. Available: \url{http://dx.doi.org/10.1007/s10515-020-00277-4}
\BIBentrySTDinterwordspacing

\bibitem{feng2024prompting}
S.~Feng and C.~Chen, ``Prompting is all you need: Automated android bug replay with large language models,'' in \emph{Proceedings of the 46th IEEE/ACM International Conference on Software Engineering}, 2024, pp. 1--13.

\bibitem{feng2020codebert}
Z.~Feng, D.~Guo, D.~Tang, N.~Duan, X.~Feng, M.~Gong, L.~Shou, B.~Qin, T.~Liu, D.~Jiang \emph{et~al.}, ``Codebert: A pre-trained model for programming and natural languages,'' \emph{arXiv preprint arXiv:2002.08155}, 2020.

\bibitem{friedman2002stochastic}
J.~H. Friedman, ``Stochastic gradient boosting,'' \emph{Computational statistics \& data analysis}, vol.~38, no.~4, pp. 367--378, 2002.

\bibitem{fu2023security}
Y.~Fu, P.~Liang, A.~Tahir, Z.~Li, M.~Shahin, and J.~Yu, ``Security weaknesses of copilot generated code in github,'' \emph{arXiv preprint arXiv:2310.02059}, 2023.

\bibitem{gao2023makes}
S.~Gao, X.-C. Wen, C.~Gao, W.~Wang, H.~Zhang, and M.~R. Lyu, ``What makes good in-context demonstrations for code intelligence tasks with llms?'' in \emph{2023 38th IEEE/ACM International Conference on Automated Software Engineering (ASE)}.\hskip 1em plus 0.5em minus 0.4em\relax IEEE, 2023, pp. 761--773.

\bibitem{geng2024large}
M.~Geng, S.~Wang, D.~Dong, H.~Wang, G.~Li, Z.~Jin, X.~Mao, and X.~Liao, ``Large language models are few-shot summarizers: Multi-intent comment generation via in-context learning,'' 2024.

\bibitem{gesi2021empirical}
J.~Gesi, J.~Li, and I.~Ahmed, ``An empirical examination of the impact of bias on just-in-time defect prediction,'' in \emph{Proceedings of the 15th ACM/IEEE international symposium on empirical software engineering and measurement (ESEM)}, 2021, pp. 1--12.

\bibitem{guo2022unixcoder}
D.~Guo, S.~Lu, N.~Duan, Y.~Wang, M.~Zhou, and J.~Yin, ``Unixcoder: Unified cross-modal pre-training for code representation,'' \emph{arXiv preprint arXiv:2203.03850}, 2022.

\bibitem{guo2024exploring}
Q.~Guo, J.~Cao, X.~Xie, S.~Liu, X.~Li, B.~Chen, and X.~Peng, ``Exploring the potential of chatgpt in automated code refinement: An empirical study,'' in \emph{Proceedings of the 46th IEEE/ACM International Conference on Software Engineering}, 2024, pp. 1--13.

\bibitem{gupta2020extracting}
\BIBentryALTinterwordspacing
A.~Gupta, B.~Suri, V.~Kumar, and P.~Jain, ``Extracting rules for vulnerabilities detection with static metrics using machine learning,'' \emph{International Journal of System Assurance Engineering and Management}, pp. 65--76, 9 2020. [Online]. Available: \url{http://dx.doi.org/10.1007/s13198-020-01036-0}
\BIBentrySTDinterwordspacing

\bibitem{harer2018automated}
J.~A. Harer, L.~Y. Kim, R.~L. Russell, O.~Ozdemir, L.~R. Kosta, A.~Rangamani, L.~H. Hamilton, G.~I. Centeno, J.~R. Key, P.~M. Ellingwood \emph{et~al.}, ``Automated software vulnerability detection with machine learning,'' \emph{arXiv preprint arXiv:1803.04497}, 2018.

\bibitem{hosseini2017systematic}
S.~Hosseini, B.~Turhan, and D.~Gunarathna, ``A systematic literature review and meta-analysis on cross project defect prediction,'' \emph{IEEE Transactions on Software Engineering}, vol.~45, no.~2, pp. 111--147, 2017.

\bibitem{husain2019codesearchnet}
H.~Husain, H.-H. Wu, T.~Gazit, M.~Allamanis, and M.~Brockschmidt, ``{CodeSearchNet} challenge: Evaluating the state of semantic code search,'' \emph{arXiv preprint arXiv:1909.09436}, 2019.

\bibitem{jiang2024surveylargelanguagemodels}
\BIBentryALTinterwordspacing
J.~Jiang, F.~Wang, J.~Shen, S.~Kim, and S.~Kim, ``A survey on large language models for code generation,'' 2024. [Online]. Available: \url{https://arxiv.org/abs/2406.00515}
\BIBentrySTDinterwordspacing

\bibitem{Jiang_2021}
\BIBentryALTinterwordspacing
N.~Jiang, T.~Lutellier, and L.~Tan, ``Cure: Code-aware neural machine translation for automatic program repair,'' in \emph{2021 IEEE/ACM 43rd International Conference on Software Engineering (ICSE)}.\hskip 1em plus 0.5em minus 0.4em\relax IEEE, May 2021. [Online]. Available: \url{http://dx.doi.org/10.1109/ICSE43902.2021.00107}
\BIBentrySTDinterwordspacing

\bibitem{katrutsa2017comprehensive}
A.~Katrutsa and V.~Strijov, ``Comprehensive study of feature selection methods to solve multicollinearity problem according to evaluation criteria,'' \emph{Expert Systems with Applications}, vol.~76, pp. 1--11, 2017.

\bibitem{khoury2023secure}
R.~Khoury, A.~R. Avila, J.~Brunelle, and B.~M. Camara, ``How secure is code generated by chatgpt?'' \emph{arXiv preprint arXiv:2304.09655}, 2023.

\bibitem{lavalley2008logistic}
M.~P. LaValley, ``Logistic regression,'' \emph{Circulation}, vol. 117, no.~18, pp. 2395--2399, 2008.

\bibitem{li2023commit}
J.~Li and I.~Ahmed, ``Commit message matters: Investigating impact and evolution of commit message quality,'' in \emph{2023 IEEE/ACM 45th International Conference on Software Engineering (ICSE)}.\hskip 1em plus 0.5em minus 0.4em\relax IEEE, 2023, pp. 806--817.

\bibitem{li2022competition}
Y.~Li, D.~Choi, J.~Chung, N.~Kushman, J.~Schrittwieser, R.~Leblond, T.~Eccles, J.~Keeling, F.~Gimeno, A.~Dal~Lago \emph{et~al.}, ``Competition-level code generation with alphacode,'' \emph{Science}, vol. 378, no. 6624, pp. 1092--1097, 2022.

\bibitem{liu2024your}
J.~Liu, C.~S. Xia, Y.~Wang, and L.~Zhang, ``Is your code generated by chatgpt really correct? rigorous evaluation of large language models for code generation,'' \emph{Advances in Neural Information Processing Systems}, vol.~36, 2024.

\bibitem{liu2023codegen4libs}
M.~Liu, T.~Yang, Y.~Lou, X.~Du, Y.~Wang, and X.~Peng, ``Codegen4libs: A two-stage approach for library-oriented code generation,'' in \emph{2023 38th IEEE/ACM International Conference on Automated Software Engineering (ASE)}.\hskip 1em plus 0.5em minus 0.4em\relax IEEE, 2023, pp. 434--445.

\bibitem{liu2023pre}
P.~Liu, W.~Yuan, J.~Fu, Z.~Jiang, H.~Hayashi, and G.~Neubig, ``Pre-train, prompt, and predict: A systematic survey of prompting methods in natural language processing,'' \emph{ACM Computing Surveys}, vol.~55, no.~9, pp. 1--35, 2023.

\bibitem{lozhkov2024starcoder2stackv2}
\BIBentryALTinterwordspacing
A.~Lozhkov, R.~Li, L.~B. Allal, F.~Cassano, J.~Lamy-Poirier, N.~Tazi, A.~Tang, D.~Pykhtar, J.~Liu, Y.~Wei, T.~Liu, M.~Tian, D.~Kocetkov, A.~Zucker, Y.~Belkada, Z.~Wang, Q.~Liu, D.~Abulkhanov, I.~Paul, Z.~Li, W.-D. Li, M.~Risdal, J.~Li, J.~Zhu, T.~Y. Zhuo, E.~Zheltonozhskii, N.~O.~O. Dade, W.~Yu, L.~Krauß, N.~Jain, Y.~Su, X.~He, M.~Dey, E.~Abati, Y.~Chai, N.~Muennighoff, X.~Tang, M.~Oblokulov, C.~Akiki, M.~Marone, C.~Mou, M.~Mishra, A.~Gu, B.~Hui, T.~Dao, A.~Zebaze, O.~Dehaene, N.~Patry, C.~Xu, J.~McAuley, H.~Hu, T.~Scholak, S.~Paquet, J.~Robinson, C.~J. Anderson, N.~Chapados, M.~Patwary, N.~Tajbakhsh, Y.~Jernite, C.~M. Ferrandis, L.~Zhang, S.~Hughes, T.~Wolf, A.~Guha, L.~von Werra, and H.~de~Vries, ``Starcoder 2 and the stack v2: The next generation,'' 2024. [Online]. Available: \url{https://arxiv.org/abs/2402.19173}
\BIBentrySTDinterwordspacing

\bibitem{lu2021codexglue}
S.~Lu, D.~Guo, S.~Ren, J.~Huang, A.~Svyatkovskiy, A.~Blanco, C.~Clement, D.~Drain, D.~Jiang, D.~Tang \emph{et~al.}, ``Codexglue: A machine learning benchmark dataset for code understanding and generation,'' \emph{arXiv preprint arXiv:2102.04664}, 2021.

\bibitem{luo2023wizardcoder}
Z.~Luo, C.~Xu, P.~Zhao, Q.~Sun, X.~Geng, W.~Hu, C.~Tao, J.~Ma, Q.~Lin, and D.~Jiang, ``Wizardcoder: Empowering code large language models with evol-instruct,'' \emph{arXiv preprint arXiv:2306.08568}, 2023.

\bibitem{mastropaolo2023robustness}
A.~Mastropaolo, L.~Pascarella, E.~Guglielmi, M.~Ciniselli, S.~Scalabrino, R.~Oliveto, and G.~Bavota, ``On the robustness of code generation techniques: An empirical study on github copilot,'' in \emph{2023 IEEE/ACM 45th International Conference on Software Engineering (ICSE)}.\hskip 1em plus 0.5em minus 0.4em\relax IEEE, 2023, pp. 2149--2160.

\bibitem{medeiros2020vulnerable}
N.~Medeiros, N.~Ivaki, P.~Costa, and M.~Vieira, ``Vulnerable code detection using software metrics and machine learning,'' \emph{IEEE Access}, 11 2020.

\bibitem{mitchell2023detectgpt}
E.~Mitchell, Y.~Lee, A.~Khazatsky, C.~D. Manning, and C.~Finn, ``Detectgpt: Zero-shot machine-generated text detection using probability curvature,'' \emph{arXiv preprint arXiv:2301.11305}, 2023.

\bibitem{nguyen2023snippet}
P.~T. Nguyen, J.~Di~Rocco, C.~Di~Sipio, R.~Rubei, D.~Di~Ruscio, and M.~Di~Penta, ``Is this snippet written by chatgpt? an empirical study with a codebert-based classifier,'' \emph{arXiv preprint arXiv:2307.09381}, 2023.

\bibitem{niu2023empirical}
C.~Niu, C.~Li, V.~Ng, D.~Chen, J.~Ge, and B.~Luo, ``An empirical comparison of pre-trained models of source code,'' 2023.

\bibitem{pan2024assessing}
W.~H. Pan, M.~J. Chok, J.~L.~S. Wong, Y.~X. Shin, Y.~S. Poon, Z.~Yang, C.~Y. Chong, D.~Lo, and M.~K. Lim, ``Assessing ai detectors in identifying ai-generated code: Implications for education,'' \emph{arXiv preprint arXiv:2401.03676}, 2024.

\bibitem{peng2023towards}
K.~Peng, L.~Ding, Q.~Zhong, L.~Shen, X.~Liu, M.~Zhang, Y.~Ouyang, and D.~Tao, ``Towards making the most of chatgpt for machine translation,'' \emph{arXiv preprint arXiv:2303.13780}, 2023.

\bibitem{peterson2009k}
L.~E. Peterson, ``K-nearest neighbor,'' \emph{Scholarpedia}, vol.~4, no.~2, p. 1883, 2009.

\bibitem{poldrack2023ai}
R.~A. Poldrack, T.~Lu, and G.~Begu{\v{s}}, ``Ai-assisted coding: Experiments with gpt-4,'' \emph{arXiv preprint arXiv:2304.13187}, 2023.

\bibitem{pradel2018deepbugs}
M.~Pradel and K.~Sen, ``Deepbugs: A learning approach to name-based bug detection,'' \emph{Proceedings of the ACM on Programming Languages}, vol.~2, no. OOPSLA, pp. 1--25, 2018.

\bibitem{radford2018improving}
A.~Radford, K.~Narasimhan, T.~Salimans, I.~Sutskever \emph{et~al.}, ``Improving language understanding by generative pre-training,'' 2018.

\bibitem{rane2024gemini}
N.~Rane, S.~Choudhary, and J.~Rane, ``Gemini or chatgpt? capability, performance, and selection of cutting-edge generative artificial intelligence (ai) in business management,'' \emph{Capability, Performance, and Selection of Cutting-Edge Generative Artificial Intelligence (AI) in Business Management (February 19, 2024)}, 2024.

\bibitem{ren2023misuse}
X.~Ren, X.~Ye, D.~Zhao, Z.~Xing, and X.~Yang, ``From misuse to mastery: Enhancing code generation with knowledge-driven ai chaining,'' in \emph{2023 38th IEEE/ACM International Conference on Automated Software Engineering (ASE)}.\hskip 1em plus 0.5em minus 0.4em\relax IEEE, 2023, pp. 976--987.

\bibitem{riedmiller2014multi}
M.~Riedmiller and A.~Lernen, ``Multi layer perceptron,'' \emph{Machine Learning Lab Special Lecture, University of Freiburg}, vol.~24, 2014.

\bibitem{rosenthal1994parametric}
R.~Rosenthal, H.~Cooper, L.~Hedges \emph{et~al.}, ``Parametric measures of effect size,'' \emph{The handbook of research synthesis}, vol. 621, no.~2, pp. 231--244, 1994.

\bibitem{ruxton2006unequal}
G.~D. Ruxton, ``The unequal variance t-test is an underused alternative to student's t-test and the mann--whitney u test,'' \emph{Behavioral Ecology}, vol.~17, no.~4, pp. 688--690, 2006.

\bibitem{shim2020deepercoder}
S.~Shim, P.~Patil, R.~R. Yadav, A.~Shinde, and V.~Devale, ``Deepercoder: Code generation using machine learning,'' in \emph{2020 10th Annual Computing and Communication Workshop and Conference (CCWC)}.\hskip 1em plus 0.5em minus 0.4em\relax IEEE, 2020, pp. 0194--0199.

\bibitem{siddiq2022securityeval}
M.~L. Siddiq and J.~C. Santos, ``Securityeval dataset: mining vulnerability examples to evaluate machine learning-based code generation techniques,'' in \emph{Proceedings of the 1st International Workshop on Mining Software Repositories Applications for Privacy and Security}, 2022, pp. 29--33.

\bibitem{strobelt2019catching}
H.~Strobelt and S.~Gehrmann, ``Catching a unicorn with gltr: A tool to detect automatically generated text,'' \emph{Catching Unicorns with GLTR}, 2019.

\bibitem{suthaharan2016decision}
S.~Suthaharan and S.~Suthaharan, ``Decision tree learning,'' \emph{Machine Learning Models and Algorithms for Big Data Classification: Thinking with Examples for Effective Learning}, pp. 237--269, 2016.

\bibitem{tufano2018deep}
\BIBentryALTinterwordspacing
M.~Tufano, C.~Watson, G.~Bavota, M.~Di~Penta, M.~White, and D.~Poshyvanyk, ``Deep learning similarities from different representations of source code,'' in \emph{Proceedings of the 15th International Conference on Mining Software Repositories}, ser. MSR '18.\hskip 1em plus 0.5em minus 0.4em\relax New York, NY, USA: Association for Computing Machinery, 2018, p. 542–553. [Online]. Available: \url{https://doi.org/10.1145/3196398.3196431}
\BIBentrySTDinterwordspacing

\bibitem{wang2017dynamic}
K.~Wang, R.~Singh, and Z.~Su, ``Dynamic neural program embedding for program repair,'' \emph{arXiv preprint arXiv:1711.07163}, 2017.

\bibitem{wang2023codet5+}
Y.~Wang, H.~Le, A.~D. Gotmare, N.~D. Bui, J.~Li, and S.~C. Hoi, ``Codet5+: Open code large language models for code understanding and generation,'' \emph{arXiv preprint arXiv:2305.07922}, 2023.

\bibitem{wang2021codet5}
Y.~Wang, W.~Wang, S.~Joty, and S.~C. Hoi, ``Codet5: Identifier-aware unified pre-trained encoder-decoder models for code understanding and generation,'' \emph{arXiv preprint arXiv:2109.00859}, 2021.

\bibitem{white2019sorting}
M.~White, M.~Tufano, M.~Martinez, M.~Monperrus, and D.~Poshyvanyk, ``Sorting and transforming program repair ingredients via deep learning code similarities,'' in \emph{2019 IEEE 26th international conference on software analysis, evolution and reengineering (SANER)}.\hskip 1em plus 0.5em minus 0.4em\relax IEEE, 2019, pp. 479--490.

\bibitem{yu2024codereval}
H.~Yu, B.~Shen, D.~Ran, J.~Zhang, Q.~Zhang, Y.~Ma, G.~Liang, Y.~Li, Q.~Wang, and T.~Xie, ``Codereval: A benchmark of pragmatic code generation with generative pre-trained models,'' in \emph{Proceedings of the 46th IEEE/ACM International Conference on Software Engineering}, 2024, pp. 1--12.

\bibitem{yu2023codeipprompt}
Z.~Yu, Y.~Wu, N.~Zhang, C.~Wang, Y.~Vorobeychik, and C.~Xiao, ``Codeipprompt: intellectual property infringement assessment of code language models,'' in \emph{International Conference on Machine Learning}.\hskip 1em plus 0.5em minus 0.4em\relax PMLR, 2023, pp. 40\,373--40\,389.

\bibitem{zhang2020retrieval}
J.~Zhang, X.~Wang, H.~Zhang, H.~Sun, and X.~Liu, ``Retrieval-based neural source code summarization,'' in \emph{Proceedings of the ACM/IEEE 42nd International Conference on Software Engineering}, 2020, pp. 1385--1397.

\bibitem{zhang2019novel}
J.~Zhang, X.~Wang, H.~Zhang, H.~Sun, K.~Wang, and X.~Liu, ``A novel neural source code representation based on abstract syntax tree,'' in \emph{2019 IEEE/ACM 41st International Conference on Software Engineering (ICSE)}.\hskip 1em plus 0.5em minus 0.4em\relax IEEE, 2019, pp. 783--794.

\bibitem{zheng2023codegeex}
Q.~Zheng, X.~Xia, X.~Zou, Y.~Dong, S.~Wang, Y.~Xue, Z.~Wang, L.~Shen, A.~Wang, Y.~Li, T.~Su, Z.~Yang, and J.~Tang, ``Codegeex: A pre-trained model for code generation with multilingual evaluations on humaneval-x,'' 2023.

\end{thebibliography}

\end{document}